\documentclass[a4paper,11pt]{article}
\pdfoutput=1

\usepackage{jcappub}
\usepackage[T1]{fontenc}
\usepackage[dvipsnames]{xcolor}
\usepackage[english]{babel}
\usepackage[utf8]{inputenc}
\usepackage{amsmath}
\usepackage{amssymb}
\usepackage{wasysym}
\usepackage{mathtools}
\usepackage[version=3]{mhchem}
\usepackage{graphicx}
\usepackage[small,bf]{caption}
\usepackage{enumitem}
\usepackage{float} 
\usepackage{tabu}
\bibliography{Sources}
\usepackage{amsthm}
\usepackage{amsfonts, graphicx, pgfplots, tikz, tikz-cd, stmaryrd}
\usepackage{graphicx}

\DeclareRobustCommand{\rchi}{{\mathpalette\irchi\relax}}
\newcommand{\irchi}[2]{\raisebox{\depth}{$#1\chi$}}

\title{Cosmic Strings from Tribrid Inflation}

\author{Stefan Antusch$^1$, Katarina Trailovi\'c$^{2,3}$}
\affiliation[1]{Department of Physics, University of Basel, Klingelbergstr.\ 82, CH-4056 Basel, Switzerland}
\affiliation[2]{Jo\v{z}ef Stefan Institute, Jamova 39, 1000 Ljubljana, Slovenia}
\affiliation[3]{Faculty of Mathematics and Physics, University of Ljubljana, Jadranska 19, 
1000 Ljubljana, Slovenia}

\emailAdd{stefan.antusch@unibas.ch}
\emailAdd{katarina.trailovic@ijs.si}

\date{\today}

\abstract{
Tribrid inflation is a class of supersymmetric inflation models where the scalar component of a matter superfield, or a $D$-flat direction of matter fields, drives inflation. Similar to Hybrid inflation, the end of inflation is reached when a ``waterfall field'', which was stabilized during inflation at a field value where the scalar potential features a large vacuum energy, starts rapidly rolling towards its minimum where a symmetry group $G$ is spontaneously broken. In contrast to standard supersymmetric Hybrid inflation, where the inflaton is a gauge singlet, in Tribrid inflation it can be a gauge non-singlet, which, via its vacuum expectation value, already breaks the gauge symmetry. This raises the question whether topological defects can still form after inflation in this class of models, and if so, which types of defects are generated. We investigate this question systematically in realisations of Tribrid inflation where $G = U(1)$ and we analyse under which conditions cosmic strings form. We find that in the considered cases where domain walls form, these are only temporary and do not invalidate the model realisations. We also discuss how our results can be used to analyse models of Tribrid inflation associated with the final step of $SO(10)$ breaking, where cosmic strings can be metastable and provide a promising explanation of the recent PTA results hinting at a stochastic gravitational wave background at nanohertz frequencies.
 }

\begin{document}
\maketitle

\newpage

\section{Introduction}

Cosmic inflation~\cite{Guth:1980zm,Albrecht:1982wi,Linde:1981mu,Linde:1983gd} is a leading paradigm for solving the horizon and flatness problems of standard Big Bang cosmology, and for explaining the origin of structure in the universe. After inflation has ended, it is followed by the epoch of reheating, during which the inflaton field responsible for inflation has to pass its energy to the Standard Model (SM) degrees of freedom. Therefore, to achieve successful reheating, the inflaton has to couple to the SM fields, which means that inflation has to be embedded into the underlying particle theory valid at energy scales where inflation takes place.

    In addition to solving the horizon and flatness problems, inflation has the further benefit that it can dilute away unwanted topological defects~\cite{Kibble:1976sj,Linde:1981mu} such as monopoles. In general, monopoles form at some stage of symmetry breaking in Grand Unified Theories (GUTs), because any GUT has to break in such a way that $U(1)_\text{em}$ is left unbroken. However, observations give strong upper bounds on the monopole abundance, which is inconsistent with the GUT predictions \cite{Preskill:1979zi}. This so-called monopole problem can be solved if the GUT breaking generates the monopoles  before inflation and if the subsequent period of accelerated expansion lasts long enough to sufficiently dilute the monopole abundance. On the other hand, topological defects such as (metastable) cosmic strings (see e.g.\ \cite{Hindmarsh:1994re} for a review) could form after inflation without causing a similar problem as the monopoles. In fact, if they form after inflation they can even leave an observable signature in the stochastic gravitational-wave background (SGWB) that can provide a valuable window into the early universe and underlying particle physics models at high energies. 

This is in particular the case in models where inflation ends via a ``waterfall'', like in hybrid inflation \cite{Linde:1993cn}, i.e.\ where a second order phase transition associated with the spontaneous breaking of a symmetry group $G$ to a subgroup $K$ terminates inflation. The phase transition is triggered by a ``waterfall field''   rapidly rolling towards the minimum of the potential, thereby destabilising the ``flat valley'' with large vacuum energy where inflation took place. Hybrid inflation can also be realized in supersymmetric models \cite{Copeland:1994vg,Dvali:1994ms,Linde:1997sj}, which provides an attractive framework for realising a sufficiently flat potential for inflation. More precisely, in supersymmetric (SUSY)  Hybrid inflation the scalar component of a gauge singlet superfield is identified with the inflaton and its $F$-term provides the vacuum energy that drives inflation.

In this work we focus on the class of {\em Tribrid inflation} \cite{Antusch:2009vg,Antusch:2008pn,Antusch:2009ef} where the scalar component of a matter superfield (e.g.\ a right-handed sneutrino, cf.\ \cite{Antusch:2004hd}), or a $D$-flat direction of matter fields \cite{Antusch:2010va,Masoud:2021prr}, is identified as the inflaton. In contrast to standard supersymmetric Hybrid inflation, where the inflaton is necessarily a gauge singlet, in Tribrid inflation it can be a gauge non-singlet, which, via its vacuum expectation value (vev), already breaks the gauge symmetry $G$ during inflation to a subgroup $K'$. This can change the dynamics of the phase transition and the symmetry breaking, and thus impact which topological defects are formed after inflation. While in standard supersymmetric Hybrid inflation the gauge group $G$ remains intact until the end of inflation and thus the topological defect production can be evaluated from the homotopy classes $\pi_n (G/K)$, the production of topological defects after inflation in Tribrid models depends, as we will show, on the specific model realisation and the resulting dynamics of the involved fields at the end of inflation.   

We will focus on the case $G = U(1)$, where we will show that even though the gauge $U(1)$ gets broken during inflation by a non-zero vev of the inflation field, after inflation cosmic strings can nevertheless get produced. We also show that in specific model classes also domain walls can form, however in all considered cases these turn out to be only temporary, and thus they do not invalidate these models. We systematically analyse various model realisations of $U(1)$ Tribrid inflation, with different inflaton-waterfall field coupling terms as well as potential ``deformation terms'' allowed by the $U(1)$ gauge symmetry. In addition, we also discuss how our results can be used to analyse Tribrid inflation embedded in $SO(10)$ GUT scenarios like the ones discussed in \cite{Antusch:2023zjk}, where the cosmic strings can be metastable and provide a promising explanation of the recent PTA results \cite{NANOGrav:2023gor,EPTA:2023fyk,Xu:2023wog,Reardon:2023gzh} hinting at a stochastic gravitational wave background at nanohertz frequencies.

The paper is organised as follows: In section 2 we will provide an overview over topological defects and Tribrid inflation, including a comparison to Hybrid inflation. In section 3 we will focus on Tribrid inflation with a gauged $U(1)$ symmetry and investigate the formation of topological defects in various scenarios. Finally, section 4 contains a discussion on possible generalisations of our results, on how $U(1)$ Tribrid inflation can be embedded in $SO(10)$ GUTs, and how our results can be applied to analyse such cases.

\section{Topological Defects in Tribrid Inflation}

Tribrid inflation \cite{Antusch:2009vg,Antusch:2008pn,Antusch:2009ef} is a variant of Hybrid inflation, where the scalar component of a matter superfield, or a $D$-flat direction of matter fields, drives inflation. These matter fields can be singlets, i.e.\ right-chiral neutrinos \cite{Antusch:2004hd}, but also gauge non-singlets \cite{Antusch:2010va,Masoud:2021prr}, e.g.\ in the context of GUTs. 

In the following, we will first compare a simple version of Tribrid inflation to standard SUSY Hybrid inflation \cite{Copeland:1994vg,Dvali:1994ms,Linde:1997sj} in order to illustrate their differences, before we turn to a more general description of the Tribrid inflation model realisations in the next section. Also, we will give an overview over topological defects and provide the tools to determine them in standard SUSY Hybrid inflation. However, as we will discuss, this formalism cannot straightforwardly be applied to the Tribrid inflation case.

\subsection{Tribrid Inflation vs. Standard SUSY Hybrid Inflation}

{\bf Standard SUSY Hybrid Inflation:}\\
In a standard SUSY Hybrid inflation model, one considers two types of superfields: an inflaton field $S$ and waterfall fields $H$, $\bar{H}$. The superpotential is typically given by 
\begin{equation}
W=\kappa S(H\bar{H}-M^2),
\end{equation}
where the chiral superfield $S$ is a singlet under a gauge group $G$, while the chiral superfields $H$ and $\bar{H}$ are in conjugate representations of each other under $G$. In the following, we will use the same letters for the scalar fields and for the corresponding superfield.

For $H=\bar{H}=0$, the $F$-term of $S$ provides the vacuum energy that drives inflation, i.e.\ $V_0=|\kappa|^2 M^4$. The squared mass eigenvalues of the waterfall fields are given by 
\begin{equation}
m^2_{1,\ldots,2r}=|\kappa|^2(|\langle S \rangle|^2+M^2) \qquad m^2_{2r+1,\ldots,4r}=|\kappa|^2(|\langle S \rangle|^2-M^2),
\end{equation}
where $r$ is the dimension of the representation of the waterfall fields. Thus, during inflation, as long as $|\langle S \rangle| > M$, the waterfall fields $H$ and $\bar{H}$ are stabilized at zero, while the singlet $S$ plays the role of the slowly rolling inflaton. Although at tree level, for unbroken SUSY and canonical K\"ahler potential, the $S$ field direction is exactly flat, a slight slope is nevertheless generated from loop effects, SUSY breaking effects and non-canonical K\"ahler potential contributions, allowing the singlet $S$ to act as a slowly rolling inflaton consistent with CMB observables (cf.\ \cite{Bastero-Gil:2006zpr,Rehman:2009nq,Nakayama:2010xf,Antusch:2012jc,Buchmuller:2014epa,Schmitz:2018nhb}).

As the inflaton rolls towards smaller field values, it reaches a critical value,
\begin{equation}
|S_{\text{crit}}|=M ,
\end{equation}
below which $2r$ of the waterfall field directions become tachyonic and are thus not stabilized at zero anymore. They quickly roll down towards the minimum of the potential with $|F_S|^2 =0$, terminating inflation in a ``waterfall'' second order phase transition.
The vevs of $H$ and $\bar{H}$ spontaneously break the gauge symmetry $G$ to some subgroup $K$. Topological defects generically form when the homotopy classes $\pi_n (G/K)$ are non-trivial. 
\\
\\
{\bf Example of Tribrid Inflation:} \\
Extending the superfield content of standard  SUSY Hybrid inflation from two types of fields, i.e.\ inflaton (singlet) and waterfall fields to three types of fields, i.e.\ singlet $S$, inflaton fields $\phi$ and $\bar{\phi}$ (non-singlets in conjugate representations of each other) and waterfall fields $H$, $\bar{H}$, one arrives at Tribrid inflation.   

An example of a superpotential where Tribrid inflation can be realized is 
\begin{equation}W=\kappa S(H\bar{H}-M^2)+\frac{\zeta}{\Lambda}(\phi \bar{\phi})(H\bar{H}).\end{equation}
The scalar component of the superfield $S$ is not the inflaton anymore, it is stabilized at zero during inflation by a large mass generated e.g.\ by SUGRA effects from a non-canonical K\"ahler potential. Its only purpose in Tribrid inflation is to provide the large vacuum energy density for driving inflation from its $F$-term.

The squared mass eigenvalues of the waterfall fields are given by 
\begin{equation}m^2_{1,\ldots,2r}=\frac{|\zeta|^2}{\Lambda^2}|\langle \phi \bar{\phi} \rangle|^2+|\kappa|^2 M^2 \qquad m^2_{2r+1,\ldots,4r}=\frac{|\zeta|^2}{\Lambda^2}|\langle \phi \bar{\phi} \rangle|^2-|\kappa|^2 M^2,\end{equation}
where $r$ again denotes the dimension of the representation of the waterfall fields.
Thus, for large enough $|\phi \bar{\phi}|$ field values, the second term in $W$ provides a positive mass term for all $H$, $\bar{H}$ directions to stabilize them at zero. Then, a $D$-flat combination of $\phi$, $\bar{\phi}$ can play the role of the inflaton, in accordance with CMB observations, since it provides a flat direction, which is only slightly lifted by loop effects, SUSY breaking effects and K\"ahler potential contributions. Similar to Hybrid inflation, when the vev of the inflaton field falls below a critical value, certain directions of $H$, $\bar{H}$ become tachyonic and the waterfall takes place which ends inflaton. 

Note that, in contrast to standard SUSY Hybrid inflation, the inflaton fields are not singlets under the gauge group $G$. 
Their non-zero vevs break the gauge group $G$ already during inflation spontaneously to a remaining group $K'$. This can affect the dynamics of the phase transition and thus topological defect formation, as we discuss in section \ref{Sec:Tribrid-U(1)} for a generalized Tribrid inflation superpotential with $G = U(1)$.

\subsection{Topological Defect Formation}
Topological defect formation after Hybrid and Tribrid inflation models are of great interest since the defects can, on the one hand, render a given model inconsistent (e.g.\ when on overabundance of monopoles is produced or when stable domain walls are generated) or, on the other hand, produce interesting potentially observable signatures, e.g.\ via a contribution to the SGWB in the case of cosmic strings or annihilating domain walls. 
\\

\noindent {\bf Topological Defects and Homotopy Groups:}
To investigate whether scalar potentials that feature spontaneous symmetry breaking allow for distinct stable topological defect solutions, the following considerations are usually made: 
When a scalar potential $V(\phi)$ with symmetry $G$ has minima at non-zero values of a scalar multiplet $\phi$, $\phi$ develops a vacuum expectation value (vev)
$\langle \phi\rangle = \phi_0$ and the symmetry is spontaneously broken. The unbroken subgroup $K$ of $G$ is the set of all elements of $G$ that leave the ground state invariant, i.e.\
\begin{equation}K=\{g\in G \ \vert \ D(g)\phi_0=\phi_0\},\end{equation}
where $D(g)$ is a representation of the element $g\in G$.
The vacuum $\phi_0$ is only one choice of the minimum of the potential and other vacua are of the form $D(g)\phi_0$ for some $g\in G$. Thus, the vacuum manifold, which is the manifold of equivalent vacua, contains the coset space 
\begin{equation}\mathcal{M}=G/K.\end{equation}

{\em Domain walls} are two-dimensional topological defects which are formed when the vacuum manifold $\mathcal{M}$ is disconnected. This is characterised by the non-triviality of the $0$-th homotopy group, i.e.\ $\mathcal{M}$ is disconnected when $\pi_0(\mathcal{M})\neq [I]$. That is because $\pi_0(\mathcal{M})$ is the set of homotopy classes of maps from $S^0=\{ \pm 1\}$ to $\mathcal{M}$, therefore one point is always mapped to a fixed base point $x\in \mathcal{M}$ and if $\mathcal{M}$ is disconnected then one can find a map $f$ where the other point of the $0$-sphere is mapped to a connected component of $\mathcal{M}$ that does not contain the base point $x$ and thus $[f]\neq [I]$. On the other hand, if $\mathcal{M}$ is connected any two maps $f,g:S^0\rightarrow \mathcal{M}$ can be continuously transformed into each other, meaning that $[f]=[g]=[I]$ and the $0$-th homotopy group would be trivial. 

{\em Cosmic strings} are one-dimensional topological defects which are formed when the vacuum manifold $\mathcal{M}$ contains unshrinkable loops. This is characterised by the non-triviality of the first homotopy group. When $\pi_1(\mathcal{M})\neq[I]$, it means that there exists a loop $f:S^1\rightarrow \mathcal{M}$ which cannot be continuously deformed to the identity map that sends all points from the circle $S^1$ to the base point $x\in \mathcal{M}$, i.e.\ $f$ is a loop which is unshrinkable to a point.
Since \begin{equation}\pi_1(U(1))\cong \mathbb{Z},\end{equation} it follows that cosmic strings can exist in a potential where an $U(1)$ symmetry gets spontaneously broken. Here one can distinguish between the cases of a global or a local $U(1)$ symmetry. 
One important difference is that from the breaking of a global $U(1)$, a massless Goldstone boson remains in the particle spectrum, which provides an additional decay mode of cosmic string loops in addition to gravitational waves. In the case of local $U(1)$ it gets eaten to generate the mass of the gauge field after symmetry breaking and this decay mode is absent.

{\em Monopoles} are point-like topological defects which are formed when the vacuum manifold $\mathcal{M}$ contains unshrinkable surfaces. This is the case when $\pi_2(\mathcal{M})\neq [I]$, because it means that there exists a map $f:S^2\rightarrow \mathcal{M}$ which cannot be continuously deformed to $I$ and thus it accords to a two-surface that cannot be shrunk to a point. 

Although the above considerations indicate that the mentioned topological defect solutions are possible, it does not mean that they are actually produced from the phase transition dynamics, and they are also not necessary conditions for topological defect production.
\\
\\
\noindent {\bf Impact of Phase Transition Dynamics on Defect Formation:}
Evaluating topological defect production solemnly from $\pi_n(\mathcal{M})$ assumes that the fields are randomly distributed on $\mathcal{M}$ by some mechanism. This is e.g.\ realized in standard SUSY Hybrid inflation, where $G$ is unbroken until the critical point is reached, followed by spontaneous symmetry breaking of $G$ to $K$ in a way that the whole vacuum manifold is reached. 
When, however, the dynamics of the phase transition is such that only a restricted region of $\mathcal{M}$ is accessible, then whether topological effects form or not cannot be evaluated from $\pi_n(\mathcal{M})$ alone, but requires including the phase transition dynamics.
Another situation where an evaluation based solemnly on $\pi_n(\mathcal{M})$ does not yield the correct result is when a symmetry is only approximate, such that the vacuum is ``deformed'', but the dynamics of the phase transition still leads to the formation of the respective topological defect. 
As we will see, such situations can occur in models of Tribrid inflation. We will therefore have to study in detail the dynamics of the phase transition.
\\

\noindent {\bf Formation of Topological Defects in Tribrid Inflation Models:} 
To evaluate whether topological defects form dynamically in models like Tribrid inflation, one has to track the evolution of the involved fields, starting from inflation. The inflaton fields are assumed to have sufficiently large initial vevs to stabilise the waterfall fields. 
We also assume that inflation lasts long enough to sufficiently dilute all defects that might have been produced before/during inflation. 
We are thus only interested in the topological defects that form after inflation.

In Tribrid inflation, as we discussed above, the masses of the waterfall fields depend on the vev of the inflaton field(s) and there can exist critical inflaton field values below which some waterfall field directions get negative squared masses, i.e.\ become tachyonic.
Before the inflaton(s) drop below the (first) critical value, all waterfall field directions have positive squared masses and are thus stabilized at the same field value everywhere in space, which implies that no topological defects associated with the waterfall fields could have formed. However, as soon as the critical value is reached, some waterfall field directions become tachyonic, which triggers the end of inflation and potentially leads to topologically non-trivial field configurations, which now are not inflated away anymore. 

To study topological defect formation one therefore has to first determine the possible critical inflaton field value(s). If none exist, then the waterfall field will end up in a unique vacuum, and no topological defects will form. As we will see, this can happen in Tribrid inflation when deformations of the potential exist due to the symmetry breaking of $G$ implied by the non-zero vev of the non-singlet inflaton field. When critical field values exist (and as we will see there can indeed exist multiple critical values), one starts with the highest one and proceeds to the lowest one.

At a critical value, one has to check whether a non-trivial field configuration of the waterfall fields $ H,\bar{H}$ can form, when the tachyonic field degrees of freedom spontaneously choose in which direction in field space they move in each region of position space. The relevant quantity to analyse is the dynamically accessible field space for the waterfall fields when $\phi, \bar \phi$ reach their critical value. When the accessible field configurations can be disconnected or have unshrinkable loops or two-surfaces, domain walls, strings or monopoles will form.

\section{Topological Defect Formation in $U(1)$ Tribrid Inflation}\label{Sec:Tribrid-U(1)}

\noindent {\bf General Form of the $U(1)$ Tribrid Inflation Superpotential:}
In this section, we investigate topological defect formation in various model realisations of Tribrid inflation with a local $U(1)$ symmetry. For definiteness, we will assume that both $\phi$ and $H$ are singly charged under the $U(1)$ gauge symmetry, such that we arrive at the following charge assignment for the involved chiral superfields $\phi,H$, their conjugates $\bar{\phi},\bar{H}$, and the singlet $S$:\footnote{Note that the charge assignment is motivated by the embedding into  $SO(10)$ GUTs described in section \ref{sec:generalisation}, but different charge assignements may also be considered.}
\begin{center}
\begin{tabular}{ c| c| c| c| c| c }
  & $S$ & $\phi$ & $\bar{\phi}$ & $H$ & $\bar{H}$\\ 
  \hline
 $U(1)$& $0$& $1$ & $-1$ & $1$ & $-1$ 
\end{tabular}.
\end{center}
The superpotential for Tribrid inflation with a local $U(1)$ symmetry will be assumed  to have the general form
\begin{equation}
    W=\kappa S (H\bar{H}-M^2)+f(H, \bar{H}, \phi, \bar{\phi})+ h (S, H, \bar{H}, \phi, \bar{\phi}),
 \label{superp}
\end{equation}
where, w.l.o.g., we can take $M$ real and positive (as this can be achieved by a global phase redefinition of $S$). 
The global SUSY scalar potential is then given by the $F$-term and the $D$-Term part
\begin{equation}V=V_F+V_D=F^{*i}F_i+\frac{1}{2} D^a D^a ,\end{equation}
where 
$D=-g \left(|\phi|^2-|\bar{\phi}|^2+|H|^2-|\bar{H}|^2\right),$ with $g$ being the gauge coupling constant, 
and $F_i=-\delta W^*/\delta \phi^{*i}$.

In order to realize Tribrid inflation, one needs a flat direction of the scalar potential with dominating vacuum energy, in which slow-roll inflation can take place for a sufficient number of e-folds, and a ``waterfall transition'' that ends the slow-roll and terminates inflation. 
Indeed, we see from the first term in Eq.~(\ref{superp}) that the $F$-term of $S$ provides the large vacuum energy density $V_0=|\kappa|^2 M^2$ which drives inflation. The $D$-flatness condition $V_D=0$ provides a field direction in which the potential is flat, given by $|\phi| = |\bar{\phi}|$ when $H = \bar H = 0$. 
During inflation, $\phi,\bar{\phi}$ have large field values satisfying $V_D=0$. We also assume that inflation lasts long enough to effectively inflate away any previously produced topological defects.
Furthermore, as already mentioned above, we will always assume that a suitable small slope of the inflaton potential is generated by loop corrections, by a small non-zero vev of $S$ during inflation, coming from the SUSY breaking sector, and/or by Planck-suppressed operators from the K\"ahler potential (cf.\ \cite{Bastero-Gil:2006zpr,Rehman:2009nq,Nakayama:2010xf,Antusch:2012jc,Buchmuller:2014epa,Schmitz:2018nhb}).

The other terms in Eq.~(\ref{superp}) have the following characteristics: $f$ contains terms that are even in the waterfall fields $H, \bar H$ and even in the inflaton fields $\phi,\bar{\phi}$. The combination of the first and second term in Eq.~(\ref{superp}) can lead to a critical point below which the waterfall takes place, as described for a specific example in the previous section. The function $h (S, H, \bar{H}, \phi, \bar{\phi})$ consists of terms which are odd in the waterfall fields, and leads to deformations of the scalar potential.
As we will see, the specific form of $f$ and $h$ affects the dynamics of the phase transition and thus also topological defect formation.
\\

\noindent {\bf Considered Cases:}
In the following, we will systematically study topological defect formation in the above-described model class. A summary of the considered cases is given in table \ref{Tab:1}. We start with taking $h =0$. We will see that there are three cases in which Tribrid inflation can be realized in a non-renormalizable theory up to dimension four operators,  with a local $U(1)$ symmetry, where in the third case $f(H, \bar{H}, \phi, \bar{\phi})$ will be the combination of the functions $f$ from the first and second case.
In the next step, we will investigate how deformations can influence the formation of topological defects. We start with terms  where $h$ is linear in the waterfall fields, and then turn to an example for a cubic deformation. In the whole deformation analysis, we will take the function $f(H, \bar{H}, \phi, \bar{\phi})$ from case 1. 
In addition, we will examine how a possible Fayet-Iliopoulos term would change the inflationary dynamics and whether it has an influence on topological defect formation.

\begin{table}[H]
\begin{center}
\begin{tabular}{ c| c| c}
  & $f$ & $h$ \\ [0.5ex]
  \hline \hline\rule{0pt}{1.05\normalbaselineskip}
 Case 1& $\frac{\zeta}{\Lambda} (\phi \bar{\phi})(H \bar{H})$ & \\[0.5ex]
  \hline\rule{0pt}{1.05\normalbaselineskip}
 Case 2 & $\frac{\lambda}{\Lambda} (\phi \bar{H})(\phi \bar{H})+\frac{\gamma}{\Lambda} (\bar{\phi} H)(\bar{\phi} H)$ &  \\[0.5ex]
  \hline\rule{0pt}{1.05\normalbaselineskip}
 Case 3  & $\frac{\zeta}{\Lambda} (\phi \bar{\phi})(H \bar{H})+\frac{\lambda}{\Lambda} (\phi \bar{H})(\phi \bar{H})+\frac{\gamma}{\Lambda} (\bar{\phi} H)(\bar{\phi} H)$ \\[0.5ex]
  \hline\rule{0pt}{1.05\normalbaselineskip}
 Case 1a  & $\frac{\zeta}{\Lambda} (\phi \bar{\phi})(H \bar{H})$& $\delta (H \bar{\phi})$\\[0.5ex]
  \hline\rule{0pt}{1.05\normalbaselineskip}
 Case 1b  & $\frac{\zeta}{\Lambda} (\phi \bar{\phi})(H \bar{H})$ & $\frac{\delta}{\Lambda} (\phi \bar{\phi})(\phi \bar{H})$\\[0.5ex]
  \hline\rule{0pt}{1.05\normalbaselineskip}
 Case 1c & $\frac{\zeta}{\Lambda} (\phi \bar{\phi})(H \bar{H})$ & $\delta S H \bar{\phi}$\\[0.5ex]
  \hline\rule{0pt}{1.05\normalbaselineskip}
 Case 1d & $\frac{\zeta}{\Lambda} (\phi \bar{\phi})(H \bar{H})$ &$\frac{\delta}{\Lambda} (\phi \bar{H})(H \bar{H})$ \\[0.5ex]
\end{tabular}
\end{center}
\caption{Overview of the cases considered in our analysis, which are defined by the choices for the terms $f$ and $g$ in the superpotential of Eq.~(\ref{superp}).
 \label{Tab:1}}
\end{table}

\subsection{Case 1}
The superpotential that we consider in the first case has the form
\begin{equation}\label{eq:W_case1}
    W=\kappa S \left( H \bar{H}- M^2\right)+\frac{\zeta}{\Lambda} (\phi \bar{\phi})(H \bar{H}),
\end{equation}
where $\Lambda$ is some cut-off scale.
\\

\noindent {\bf Model Realization:} 
Before we turn to the analysis of the phase transition dynamics, let us discuss how one can arrive at this superpotential from an underlying symmetry. For this purpose, we introduce an additional chiral superfield $X$, which we will assume to acquire some large vev $\langle X \rangle$ from an unspecified further sector of the model, and consider e.g.\ the following charge assignments for the fields under the listed symmetries,
\begin{center}
\begin{tabular}{ c| c| c| c| c| c | c}
  & $S$ & $\phi$ & $\bar{\phi}$ & $H$ & $\bar{H}$ & $X$\\ 
  \hline \hline
 $U(1)$& $0$& $1$ & $-1$ & $1$ & $-1$& $0$\\
 \hline
 $U(1)_R$& $1$ & $1/2$ & $1/2$ & $0$ & $0$ & $0$\\
 \hline
 $\mathbb{Z}_4$ & $0$ & $0$ & $1$ & $2$ & $1$ &$1$
\end{tabular}
\end{center}
where, in addition to the local $U(1)$ symmetry, we have considered a $U(1)_R$-symmetry and a discrete $\mathbb{Z}_4$ symmetry. With this superfield content and these symmetries, the non-renormalizable superpotential, up to dimension four operators, is given by
\begin{equation}W=\kappa S \left(\frac{X}{\Lambda} H \bar{H}- M^2\right)+\frac{\zeta}{\Lambda} (\phi \bar{\phi})(H \bar{H}).\end{equation}
Note that the superfield $X$ and the $\mathbb{Z}_4$ symmetry have been introduced in order to forbid a direct mass term for the inflaton fields $\phi$ and $\bar{\phi}$, which, when too large, would spoil the flatness of the inflaton potential, and at the same time to allow the term $(\phi \bar{\phi})(H \bar{H})$. This cannot be achieved without the $X$ field using only a $\mathbb{Z}_n$ symmetry for some $n \in \mathbb{N}$, because to forbid the mass term for the inflaton fields, $(\phi \bar{\phi})$ must have a non-zero charge under $\mathbb{Z}_n$, but then to allow $(\phi \bar{\phi})(H \bar{H})$, $(H \bar{H})$ must also have a non-zero charge and then the term $S H \bar{H}$ would be forbidden since $S$ must have zero charge under $\mathbb{Z}_n$ to allow the term $-M^2 S$.
Using the superfield $X$ and the $\mathbb{Z}_4$ symmetry,  the term $S H \bar{H}$ would be still forbidden, however now one can generate the term $\frac{X}{\Lambda} S H \bar{H}$ which is allowed by the symmetries. After the gauge singlet field $X$ gets a vev, it breaks the discrete symmetry and dynamically generates the desired term $\frac{\langle X \rangle}{\Lambda} S H \bar{H}$. In the following, we will absorb the vev of $X$ into the constant $\kappa$, which then results in the superpotential of Eq.~(\ref{eq:W_case1}).
\\

\noindent  {\bf Phase Transition Dynamics:} The scalar potential derived from Eq.~(\ref{eq:W_case1}) is given by
$$V=\left|\kappa \left( H \bar{H}- M^2\right)\right|^2+\left|\frac{\zeta}{\Lambda} \bar{\phi}(H \bar{H})\right|^2+\left|\frac{\zeta}{\Lambda} \phi (H \bar{H})\right|^2+\left|\kappa S \bar{H} +\frac{\zeta}{\Lambda} (\phi \bar{\phi}) \bar{H}\right|^2$$
\begin{equation}+\left|\kappa S H +\frac{\zeta}{\Lambda} (\phi \bar{\phi}) H\right|^2
+\frac{g^2}{2} \left(|\phi|^2-|\bar{\phi}|^2+|H|^2-|\bar{H}|^2\right)^2,
\label{scalarpot1}
\end{equation}
where we use the same symbol for the scalar fields as for the superfields.
We first assume that the gauge singlet scalar field $S$ has a zero vev, both during and after inflation. This can be achieved via non-canonical terms in the Kähler potential, which can provide a mass lager than the Hubble scale for $S$ (see e.g.\ \cite{Antusch:2012jc}).

During inflation the waterfall fields have positive squared-masses (as we will calculate below) and thus their vevs are zero at this stage, i.e.\ $\langle H \rangle =\langle \bar{H} \rangle=0$. The inflaton potential then becomes
\begin{equation}V_{\text{inf}}=|\kappa|^2 M^4 +\frac{g^2}{2} \left(|\phi|^2-|\bar{\phi}|^2\right)^2 + \Delta V .\end{equation}
We see that the $D$-term forces the inflaton fields into a $D$-flat direction where $|\langle\phi\rangle|=|\langle\bar{\phi}\rangle|$ during inflation. Thus, we get a large positive vacuum energy density $V_0=|\kappa|^2 M^4$ that drives inflation along the $D$-flat direction. Due to the $D$-flatness condition, we effectively have a one-field inflation model. 
As already mentioned earlier, the leading order inflaton potential gets additional contributions $\Delta V$ from the Kähler potential, from SUSY breaking effects as well as by loop corrections. We assume that this provides the slope for slow-roll inflation where the inflaton moves towards smaller field values. Inflation ends as soon as the waterfall fields become tachyonic and the waterfall happens. 

To derive the critical inflaton field value at which the waterfall starts, we calculate the squared masses of the waterfall fields during inflation. Plugging the inflaton vevs into Eq.~(\ref{scalarpot1}) and expanding around $\langle H \rangle =\langle \bar{H} \rangle=0$ in the basis $\mathcal{H}=\left( H \ H^* \ \bar{H} \ \bar{H}^* \right)^{T}$, we obtain the squared mass matrix $\mathbf{m}_\mathcal{H}^2$ defined by 
 $   V=\frac{1}{2} \mathcal{H}^{\dag} \mathbf{m}_\mathcal{H}^2 \mathcal{H}+ \ldots$,
where the dots denote all terms that do not contribute to the mass term of the waterfall fields. We find that the eigenvalues of $\mathbf{m}_\mathcal{H}^2$ are
\begin{equation}
m_{1,2}^2=\frac{|\zeta|^2}{\Lambda^2} |\langle \phi \rangle|^4+|\kappa|^2 M^2, \quad m_{3,4}^2=\frac{|\zeta|^2}{\Lambda^2} |\langle \phi \rangle|^4-|\kappa|^2 M^2,
\end{equation}
with the corresponding eigenstates
\begin{equation}v_1=-H+\bar{H}^*,\quad v_2=-H^*+\bar{H}, \quad v_3= H+\bar{H}^*, \quad v_4=H^*+\bar{H}.\end{equation}
Since $m_{1,2}^2 > 0$ for any $\langle \phi \rangle$, we always have that $\langle v_1 \rangle=\langle v_2 \rangle =0$ and thus $\langle \bar{H}\rangle=\langle H^*\rangle$.\footnote{This condition also holds true after symmetry breaking. Finally, the scalar fields will settle in the SUSY conserving global minimum, where $\langle \phi \rangle=\langle \bar{\phi} \rangle=0$, $\langle H \rangle=M e^{i\theta(\mathbf{x})}, \; \langle \bar{H} \rangle=M e^{-i\theta(\mathbf{x})}$ and where the angle $\theta(\mathbf{x})$ can have different values for different points $\mathbf{x}$ in space.
} 
From $m_{3,4}^2$, we obtain that the critical value, below which the waterfall fields become tachyonic, is given by
\begin{equation}|\phi_{\text{crit}}|=\sqrt{\frac{|\kappa| M}{|\zeta|}\Lambda}.\end{equation}
\\

\noindent  {\bf Topological Defect Formation:} 
From the above considerations, we know that at the critical point $\text{arg}(\langle \bar{H}\rangle )\equiv-\text{arg}(\langle H\rangle ) (\text{mod } 2\pi)$ and $|\langle \bar{H}\rangle|=|\langle H\rangle|$. Plugging this into the potential, we see that the potential $V$ only depends on $| H |$ and is independent of $\text{arg}(H)$.
As soon as the waterfall fields become tachyonic, the modulus of $\langle H\rangle$ will quickly move towards the non-zero value $|H|=M$ that minimizes the potential, while the phase $\text{arg}(\langle H\rangle)$ takes a random value $\text{arg}(\langle H\rangle)\in [0,2\pi)$. Since there is no preferred value for $\text{arg}(\langle H\rangle)$, it can happen that if we make a circle in space we go through all values from $0$ to $2 \pi$ (one or several times, i.e.\ several windings) and thus,  due to the continuity of the field, inside the circle there must be a point where $\langle H \rangle =0$. When this happens, a cosmic string has formed.
\\

\noindent  {\bf Remark on the Gauge Boson Mass:} The local $U(1)$ symmetry gets broken during inflation and therefore we have a would-be Goldstone boson that gets eaten up by the gauge boson of the $U(1)$. Since the Lagrangian contains the terms 
$$\mathcal{L}=(D_\mu \phi)(D^\mu \phi)^\dag + (D_\mu \bar{\phi})(D^\mu \bar{\phi})^\dag + (D_\mu H)(D^\mu H)^\dag + (D_\mu \bar{H})(D^\mu \bar{H})^\dag+\ldots$$
\begin{equation}=\frac{1}{2} \left[2 g^2  \left(|\langle\phi\rangle|^2+|\langle\bar{\phi}\rangle|^2+|\langle H\rangle|^2+|\langle \bar{H}\rangle|^2\right)\right] A_\mu A^\mu + \ldots,\end{equation}
we see that the gauge boson $A_\mu$ has a non-zero squared mass $m_A^2=4 g^2 |\langle \phi \rangle|^2$ from the inflaton vevs during inflation. After inflation,  $|\langle \phi \rangle|$ goes to zero but $|\langle H \rangle|$ becomes non-zero, such that in the global minimum we get $m_A^2= 4 g^2  |\langle H\rangle|^2 = 4 g^2 M^2$. Since SUSY is reestablished in the global minimum, the gaugino has the same mass and we have a would-be Goldstone superfield from the breaking of the gauged $U(1)$, which is given by \begin{equation}G=H+(-e^{-2i\theta})H^*+(-e^{-2i\theta})\bar{H}+\bar{H}^*.\end{equation}
This would-be Goldstone superfield is eaten up by the gauge superfield of the local $U(1)$.

\subsection{Case 2}
As the second case, we consider the following superpotential:
\begin{equation}W=\kappa S \left( H \bar{H}- M^2\right)+\frac{\lambda}{\Lambda} (\phi \bar{H})(\phi \bar{H})+\frac{\gamma}{\Lambda} (\bar{\phi} H)(\bar{\phi} H),\label{superp2}\end{equation}
where, again, $\Lambda$ denotes some cut-off scale.
\\

\noindent {\bf Model Realization:} For the model realisation of this case, we may consider for instance the following chiral superfields and symmetries
\begin{center}
\begin{tabular}{ c| c| c| c| c| c }
  & $S$ & $\phi$ & $\bar{\phi}$ & $H$ & $\bar{H}$ \\ 
  \hline \hline
 $U(1)$& $0$& $1$ & $-1$ & $1$ & $-1$\\
 \hline
 $U(1)_R$& $1$ & $1/2$ & $1/2$ & $0$ & $0$\\
 \hline
 $\mathbb{Z}_2$ & $0$ & $0$ & $1$ & $0$ & $0$
\end{tabular} 
\end{center}
where, in addition to the local $U(1)$ and the $U(1)_R$ symmetry, we have introduced a $\mathbb{Z}_2$ symmetry. With this superfield content and these symmetries, the non-renormalizable superpotential, up to dimension four operators, is exactly the superpotential given in Eq.~(\ref{superp2})
The purpose of the $\mathbb{Z}_2$ symmetry is to forbid a direct mass term for the inflaton fields, to preserve the flatness of the inflaton potential. Note that in this case, compared to case 1, we do not need to add a gauge singlet field $X$, since we do not require the term $\phi \bar{\phi} H \bar{H}$.
\\

\noindent  {\bf Phase Transition Dynamics:} The scalar potential derived from Eq.~(\ref{superp2}) is given by
$$V=\left|\kappa \left( H \bar{H}- M^2\right)\right|^2+\left|2\frac{\lambda}{\Lambda} (\phi \bar{H}) \bar{H}\right|^2+\left|2\frac{\gamma}{\Lambda} (\bar{\phi} H) H \right|^2+\left|\kappa S \bar{H} +2\frac{\gamma}{\Lambda} (\bar{\phi} H) \bar{\phi}\right|^2$$\begin{equation}+\left|\kappa S H +2\frac{\lambda}{\Lambda} (\phi \bar{H}) 
\phi\right|^2
+\frac{g^2}{2} \left(|\phi|^2-|\bar{\phi}|^2+|H|^2-|\bar{H}|^2\right)^2. \label{scalarpot2}
\end{equation}
Again, we assume that the singlet field gets a large mass from the Kähler potential such that it is stabilized at zero. During inflation the waterfall fields are also stabilized at zero, since they have positive squared masses. Furthermore, the $D$-term forces the inflaton fields into the $D$-flat direction where $ |\langle \phi \rangle| = |\langle \bar{\phi}\rangle| $. 
Calculating the eigenvalues of $\mathbf{m}_\mathcal{H}^2$ during inflation we obtain
\begin{equation}m_{1,2}^2=2\left(\frac{|\lambda|^2}{\Lambda^2}+\frac{|\gamma|^2}{\Lambda^2}\right) |\langle \phi \rangle|^4+\sqrt{4 \left(\frac{|\lambda|^2}{\Lambda^2}-\frac{|\gamma|^2}{\Lambda^2}\right)^2 |\langle \phi \rangle|^8+M^4|\kappa|^4} ,\end{equation}
\begin{equation}m_{3,4}^2=2\left(\frac{|\lambda|^2}{\Lambda^2}+\frac{|\gamma|^2}{\Lambda^2}\right) |\langle \phi \rangle|^4-\sqrt{4 \left(\frac{|\lambda|^2}{\Lambda^2}-\frac{|\gamma|^2}{\Lambda^2}\right)^2 |\langle \phi \rangle|^8+M^4|\kappa|^4} .\end{equation}
We see that $m_{1,2}^2>0$ for any $\langle \phi \rangle$, and that the critical inflaton field value, below which $m^2_{3,4}<0$, is given by
\begin{equation}| \phi_{\text{crit}}|=\left(\frac{M^2 |\kappa|^2}{4|\gamma| |\lambda|} \Lambda^2\right)^{1/4} .
\end{equation}
The eigenstates of $\mathbf{m}_\mathcal{H}^2$ are given by
\begin{equation}v_1=\alpha_- H +\bar{H}^*, \quad v_2=\alpha_- H^* +\bar{H}, \quad v_3=\alpha_+ H +\bar{H}^*, \quad v_4=\alpha_+ H^* +\bar{H},\end{equation}
where
\begin{equation}
\alpha_{\pm}=\frac{2 \left(\frac{|\lambda|^2}{\Lambda^2}-\frac{|\gamma|^2}{\Lambda^2}\right)|\langle \phi \rangle|^4\pm \sqrt{4 \left(\frac{|\lambda|^2}{\Lambda^2}-\frac{|\gamma|^2}{\Lambda^2}\right)^2|\langle \phi \rangle|^8+M^4|\kappa|^4}}{M^2|\kappa|^2} .
\end{equation}
We note that having only one of the two non-renormalizable terms, i.e.\ $\gamma=0$ and $\lambda\neq0$ or $\gamma\neq0$ and $\lambda=0$, does not allow for inflation. This is because two waterfall field directions would then be negative for any value of $|\phi|$ and thus there would not exist a stable flat valley where inflation can take place. 
\\

\noindent  {\bf Topological Defect Formation:} Since $m_{1,2}^2 > 0$ for any $\langle \phi \rangle$,$\langle \bar{\phi} \rangle$, we have that $\langle v_1 \rangle=\langle v_2 \rangle =0$ and thus $\langle \bar{H}\rangle=(-\alpha_-)\langle H^*\rangle$. This means that at the critical point $|\langle \bar{H}\rangle|=(-\alpha_-)|\langle H\rangle|$ (note  $-\alpha_- >0$) and $\text{arg}(\langle \bar{H}\rangle )\equiv-\text{arg}(\langle H\rangle) (\text{mod } 2\pi) $. Plugging this into the potential, it follows that the potential only depends on $|H|$ and is independent of $\text{arg}(H)$. So we see that as soon as the waterfall fields become tachyonic, there is a degeneracy in $\text{arg}(\langle \bar{H}\rangle)$ and $|H|$ quickly moves towards the minimum where $|H|=M$. Consequently, as in case 1, cosmic strings form at the critical point.

\subsection{Case 3}
Case 3 combines the terms of the first two cases to the following superpotential:
\begin{equation}W=\kappa S \left( H \bar{H}- M^2\right)+\frac{\zeta}{\Lambda} (\phi \bar{\phi})(H \bar{H})+\frac{\lambda}{\Lambda} (\phi \bar{H})(\phi \bar{H})+\frac{\gamma}{\Lambda} (\bar{\phi} H)(\bar{\phi} H).\label{superp3}
\end{equation}
Again, $\Lambda$ is some cut-off scale.
\\

\noindent {\bf Model Realization:} We consider e.g.\ the following chiral superfields and symmetries
\begin{center}
\begin{tabular}{ c| c| c| c| c| c | c}
  & $S$ & $\phi$ & $\bar{\phi}$ & $H$ & $\bar{H}$ & $X$\\ 
  \hline \hline
 $U(1)$& $0$& $1$ & $-1$ & $1$ & $-1$& $0$\\
 \hline
 $U(1)_R$& $1$ & $1/2$ & $1/2$ & $0$ & $0$ & $0$\\
 \hline
 $\mathbb{Z}_4$ & $0$ & $1$ & $0$ & $2$ & $1$ &$1$ 
\end{tabular}
\end{center}
where in addition to the local $U(1)$ and the $U(1)_R$ symmetry, we have introduced a $\mathbb{Z}_4$ symmetry. 
This is similar to case 1, apart from the swapped $\mathbb{Z}_4$-charges of $\phi,\bar{\phi}$.  
With this superfield content and symmetries, the non-renormalizable superpotential, up to dimension four operators, is given by
\begin{equation}
W=\kappa S \left(\frac{X}{\Lambda} H \bar{H}- M^2\right)+\frac{\zeta}{\Lambda} (\phi \bar{\phi})(H \bar{H})+\frac{\lambda}{\Lambda} (\phi \bar{H})(\phi \bar{H})+\frac{\gamma}{\Lambda} (\bar{\phi} H)(\bar{\phi} H).
\end{equation}
The gauge singlet field $X$ and the $\mathbb{Z}_4$ symmetry were introduced for the same reasons as in case 1. Again, we assume that $X$ gets a vev from some additional sector of the theory and absorb it into the constant $\kappa$. This yields the desired superpotential of Eq.~(\ref{superp3}).
\\

\noindent  {\bf Phase Transition Dynamics:} The scalar potential derived from Eq.~(\ref{superp3}) is given by
$$V=\left|\kappa \left( H \bar{H}- M^2\right)\right|^2+\left|\frac{\zeta}{\Lambda} \bar{\phi}(H \bar{H})+2\frac{\lambda}{\Lambda} (\phi \bar{H}) \bar{H}\right|^2+\left|\frac{\zeta}{\Lambda} \phi (H \bar{H})+2\frac{\gamma}{\Lambda} (\bar{\phi} H) H\right|^2$$$$+\left|\kappa S \bar{H} +\frac{\zeta}{\Lambda} (\phi \bar{\phi}) \bar{H}+2\frac{\gamma}{\Lambda} (\bar{\phi} H) \bar{\phi}\right|^2+\left|\kappa S H +\frac{\zeta}{\Lambda} (\phi \bar{\phi}) H+2\frac{\lambda}{\Lambda} (\phi \bar{H}) 
\phi\right|^2$$\begin{equation}
+\frac{g^2}{2} \left(|\phi|^2-|\bar{\phi}|^2+|H|^2-|\bar{H}|^2\right)^2.\label{scalarpot3}
\end{equation}
We assume that the singlet field $S$ is stabilized at zero during and after inflation. For large enough inflaton vevs, the waterfall fields have positive squared masses during inflation and thus zero vevs. Also, the $D$-term forces the inflaton fields into a $D$-flat direction, where $ |\langle \phi \rangle| = |\langle \bar{\phi}\rangle| $ and thus the eigenvalues of $\mathbf{m}^2_\mathcal{H}$ during inflation are 

 \begin{eqnarray}
m^2_{1,2,3,4}&=&\left(2 \frac{|\gamma|^2}{\Lambda^2}+ 2\frac{|\lambda|^2}{\Lambda^2}+\frac{|\zeta|^2}{\Lambda^2}\right)|\phi|^4 \nonumber\\ 
&&  \pm \sqrt{4\left(\frac{|\gamma|^2}{\Lambda^2}- \frac{|\lambda|^2}{\Lambda^2}\right)^2 |\phi|^8 + \left( 2 \frac{|\zeta \gamma^*+\lambda \zeta^*|}{\Lambda^2}|\phi|^4\pm M^2|\kappa|^2\right)^2} ,
\end{eqnarray}

\noindent with $m^2_1 \sim (+,+), \quad m^2_2\sim (+,-), \quad m^2_3\sim (-,+), \quad m^2_4\sim (-,-)$, where the signs in the bracket correspond to the choices of the signs in the formula above.
In this case, the choice of parameters decides whether inflation and a waterfall transition can occur. We can do a redefinition of the fields in order to make all parameters real and positive. Then, any choice of parameters, except the choice that fulfills 
\begin{equation}
    4\gamma \lambda
=\zeta^2, \label{eq2}
\end{equation}
lead to a waterfall transition. For the choice of parameters that fulfills Eq.~(\ref{eq2}) we get $m_3^2<0$ for any $|\phi|$ and thus in that case there would be no inflation. 

Since the eigenvalues $m^2_{1,2}>0$ for any value of the inflaton, these directions are always stabilized at zero. On the other hand, $m^2_3$ and $m^2_4$ can get tachyonic, where $m_3^2<0$ for
\begin{equation}
|\phi|<|\phi_{\text{crit}1}|=\left(2 M^2 |\kappa|^2\frac{2 \frac{|\zeta \gamma^*+\lambda \zeta^*|}{\Lambda^2}+\sqrt{\frac{|\zeta|^4}{\Lambda^4}+4\left(\frac{|\gamma|^2}{\Lambda^2}+\frac{|\lambda|^2}{\Lambda^2}\right)\frac{|\zeta|^2}{\Lambda^2}+16\frac{|\gamma|^2|\lambda|^2}{\Lambda^4}}}{\frac{|\zeta|^4}{\Lambda^4}+4\left(\frac{|\gamma|^2}{\Lambda^2}+\frac{|\lambda|^2}{\Lambda^2}\right)\frac{|\zeta|^2}{\Lambda^2}+16\frac{|\gamma|^2|\lambda|^2}{\Lambda^4}}\right)^{1/4}\end{equation}
and $m_4^2<0$ for
\begin{equation}|\phi|<|\phi_{\text{crit}2}|=\left(2 M^2 |\kappa|^2\frac{-2 \frac{|\zeta \gamma^*+\lambda \zeta^*|}{\Lambda^2}+\sqrt{\frac{|\zeta|^4}{\Lambda^4}+4\left(\frac{|\gamma|^2}{\Lambda^2}+\frac{|\lambda|^2}{\Lambda^2}\right)\frac{|\zeta|^2}{\Lambda^2}+16\frac{|\gamma|^2|\lambda|^2}{\Lambda^4}}}{\frac{|\zeta|^4}{\Lambda^4}+4\left(\frac{|\gamma|^2}{\Lambda^2}+\frac{|\lambda|^2}{\Lambda^2}\right)\frac{|\zeta|^2}{\Lambda^2}+16\frac{|\gamma|^2|\lambda|^2}{\Lambda^4}}\right)^{1/4}.
\end{equation}
We see that $|\phi_{\text{crit}1}|=|\phi_{\text{crit}2}|$ only for the choices $\zeta\neq 0$ and $\gamma=\lambda=0$ or $\zeta=0$ and $\gamma\neq0$ and $\lambda\neq0$, where we used again the assumption that all parameters are real and positive (after a possible field redefinition) and that  
condition (\ref{eq2}) is not fulfilled. These are exactly the cases 1 and 2 that we investigated before. Any other choice of parameters that does not fulfill condition (\ref{eq2}) gives 
$|\phi_{\text{crit}1}|>|\phi_{\text{crit}2}|$ and thus $m_1^2$ becomes tachyonic first. The fact that we now have two critical points, compared to only one critical point for the cases 1 and 2, significantly changes the phase transition dynamics, as we now discuss.
\\

\noindent  {\bf Topological Defect Formation:}
When 
$|\phi|>|\phi_{\text{crit}1}|$, both waterfall fields are stabilized at zero. As soon as $|\phi_{\text{crit}2}|<|\langle \phi \rangle|
<|\phi_{\text{crit}1}|$ one of the waterfall field direction gets destabilized. 
The eigenstates corresponding to the eigenvalues 
$m^2_{1,2,4}>0$ still have zero vevs, which can be shown to imply the conditions 
\begin{equation}
\langle H\rangle =\pm |\langle H \rangle| e^{i\frac{\varphi}{2}}, \quad\langle \bar{H}\rangle =\mp  \alpha |\langle H \rangle| e^{-i\frac{\varphi}{2}},
\end{equation}
where 
\begin{equation}
\varphi=\text{arg}\left[\frac{\bar{\phi}}{\phi} \left(\frac{\zeta^* \gamma+\zeta \lambda^*}{\Lambda^2}\right)\right]+\pi\end{equation}\begin{equation} \alpha=-\frac{2 |\phi|^4\left(\frac{|\gamma|^2}{\Lambda^2}-\frac{|\lambda|^2}{\Lambda^2}\right)+ \sqrt{4 |\phi|^8 \left(\frac{|\gamma|^2}{\Lambda^2}-\frac{|\lambda|^2}{\Lambda^2}\right)^2+\left(2\frac{|\zeta \gamma^*+\lambda \zeta^*|}{\Lambda^2}|\phi|^4+M^2|\kappa|^2\right)^2}}{2\frac{|\zeta \gamma^*+\lambda \zeta^*|}{\Lambda^2}|\phi|^4+M^2|\kappa|^2}.
\end{equation}
We see that when the modulus of the inflaton field drops below the first critical value and one waterfall field direction corresponding to $m_3^2$ becomes tachyonic, there are two possible choices for the vev and consequently domain walls form.

On these domain walls we still have $\langle H \rangle=\langle \bar{H} \rangle=0$. When $|\langle \phi \rangle|$ drops below the second critical value $|\phi_{\text{crit}2}|$, now also $m_4^2<0$ (while still $m^2_{1,2}>0$), and the corresponding waterfall field direction has the choice to roll to one of the two possible nonzero vevs. This leads to the formation of cosmic strings on top of the domain walls. 

To see this, we consider that for the waterfall field eigenvectors $\langle v_3\rangle \neq 0$, $\langle v_4\rangle \neq 0$ and $\langle v_1\rangle=\langle v_2\rangle=0$ holds, from which it follows that
\begin{equation}\langle \bar{H} \rangle=\frac{1}{2} \left[(\beta-\alpha) e^{- i \varphi} \langle H \rangle -(\beta+\alpha) \langle H^*\rangle\right] \end{equation} 
with 
\begin{equation}
\beta= -\frac{2 |\phi|^4\left(\frac{|\gamma|^2}{\Lambda^2}-\frac{|\lambda|^2}{\Lambda^2}\right)+ \sqrt{4 |\phi|^8 \left(\frac{|\gamma|^2}{\Lambda^2}-\frac{|\lambda|^2}{\Lambda^2}\right)^2+\left(2\frac{|\zeta \gamma^*+\lambda \zeta^*|}{\Lambda^2}|\phi|^4-M^2|\kappa|^2\right)^2}}{-2\frac{|\zeta \gamma^*+\lambda \zeta^*|}{\Lambda^2}|\phi|^4+M^2|\kappa|^2}.
\end{equation}
So the only free parameters are the modulus of the vev of $H$ and its phase. Below the second critical point, starting from $\langle \bar{H} \rangle = 0$ on top of the domain wall, $H$ can fall into any direction in the complex plane, i.e.\ $\text{arg}(H)$ can take any value. Thus, cosmic strings form on top of the domain wall. 

\begin{figure}[h]
    \centering
    \includegraphics[width=1\textwidth]{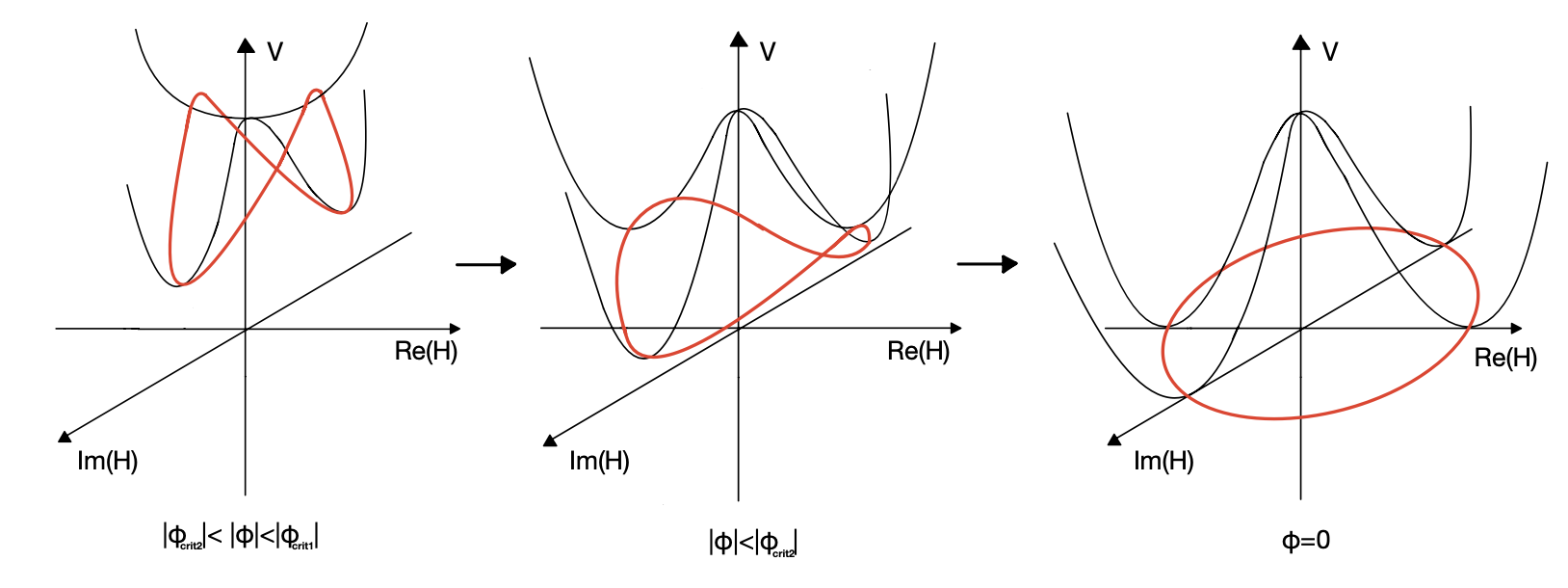}
    \caption{Illustration of the evolution of the potential of $H$ as the inflaton rolls towards zero. On the left: the potential of $H$ when $|\phi_{\text{crit}2}|<|\langle \phi\rangle|<|\phi_{\text{crit}1}|$, where domain walls have formed. In the middle: the potential of $H$ when $|\langle \phi\rangle|<|\phi_{\text{crit}2}|$, where cosmic strings have formed on top of the domain walls, which now have lower energy density. On the right: the potential of $H$ when $|\langle \phi\rangle|=|\langle \bar{\phi}\rangle|=0$, where the domain walls have vanished while the cosmic strings still persist. }
    \label{potfig}
\end{figure}

We note that the domain walls do not fully disappear as long as $|\langle \phi\rangle|,|\langle \bar{\phi}\rangle|$ are nonzero and lift the degeneracy of the waterfall potential. They remain as defects with lower energy. 
As $|\langle \phi\rangle|,|\langle \bar{\phi}\rangle|$ decrease, the domain walls become energetically lower until they disappear completely when $|\langle \phi\rangle|=|\langle \bar{\phi}\rangle|=0$ and the potential becomes again fully degenerate in $\text{arg}(H)$.  
Finally, when the global minimum is reached, only the cosmic strings remain and 
$|\langle H\rangle |=|\langle \bar{H}\rangle|=M$, $\text{arg}(\langle H\rangle )\equiv-\text{arg}(\langle \bar{H}\rangle) \ (\text{mod } 2 \pi)$.
An illustration is given in Fig. \ref{potfig}.

\subsection{Comment on the Failure of Symmetry Arguments}   \label{sec:symmargfailure}
We have seen in the previous subsections that despite the $U(1)$ gauge symmetry already being broken by the vev of the inflaton, cosmic strings (plus in case 3 temporary domain walls) nevertheless form during the waterfall that ends inflation. 
One might think that one can now judge which topological defects get produced by considering the homotopy groups of $K'/K$, with $K'$ being the subgroup to which the vevs of $\phi,\bar\phi$ break $G$ during inflation. However, as we now discuss, while this would give the correct answer for the cases 1 and 2, it fails for case 3.

In fact, the scalar potential of case 3, given in Eq.~(\ref{scalarpot3}), has no global $U(1)$ symmetry unbroken. It only has a $\mathbb{Z}_2$ symmetry 
\begin{center}
\begin{tabular}{c | c| c }
  & $H$ & $\bar{H}$\\ 
  \hline
 $\mathbb{Z}_2$& $1$ & $1$ 
\end{tabular},
\end{center}
which gets spontaneously broken at the critical point(s). Based on this consideration one might think that domain walls form but no cosmic strings. As we discussed above, this conclusion would be incorrect. To arrive at the correct answer a careful consideration of the waterfall dynamics at the two critical points is necessary.

\subsection{Cases 1, 2, 3 with $\langle S \rangle \neq 0$}

Let us, for completeness, also comment on the cases 1, 2, 3 with $\langle S \rangle \neq 0$ (but $| \langle S \rangle | \ll M$). Such a non-zero vev of the $S$ field can e.g.\ be induced by a linear term in the scalar potential generated from the SUSY breaking sector (see e.g.\ \cite{Buchmuller:2014epa}).

Considering $\langle S \rangle \neq 0$ for case 1, we find that it does not qualitatively change the results for topological defect formation. This is because, although the squared masses of the waterfall field directions change due to the non-zero $S$ vev, we still have
$m_1^2=m_2^2$ and $m_3^2=m_4^2$,
where $m_{1,2}^2$ are always positive and $m_{3,4}^2$ become negative as soon as the critical value is reached (which now also depends on the value of $S$). This means that there is again only one critical value and at this value two waterfall directions become tachyonic simultaneously. Hence, again cosmic strings form.

Also in case 3, a non-zero $S$ value does not qualitatively affect topological defect formation, since all four waterfall field squared masses are again non-degenerate, with two of them always being positive and the other two becoming negative at different critical values. The result is again the formation of temporary domain walls and cosmic strings.

For case 2, however, $\langle S \rangle \neq 0$ leads to a qualitatively different scenario, since now the four eigenvalues of the waterfall field squared masses are all non-degenerate. Two of them are always positive
and two get negative at different critical values. Therefore, topological defect formation happens as in case 3, i.e.\ case 2 with $\langle S \rangle \neq 0$ now also leads to the formation of temporary domain walls as well as cosmic strings forming on top of them.

\subsection{Deformations of the Waterfall Potential}

Another feature of Tribrid inflation is the possibility of deformations of the waterfall field potential from terms $h (S, H, \bar{H}, \phi, \bar{\phi})$ in Eq.~(\ref{superp}), i.e.\ terms that involve the inflaton fields and that are odd in waterfall fields. With nonzero $\langle \phi\rangle$ and $\langle \bar{\phi}\rangle$ during inflation they can induce deformations of the potential and modify the waterfall dynamics such that topological defect formation is affected. As we will see, the presence of such deformation terms can effectively shut down topological defect production. A summary of the examples for $h$ which we will discuss in the following subsections can be found in table \ref{Tab:1}.

For the discussion of deformation terms we will focus on $f(H, \bar{H}, \phi, \bar{\phi})$ of case 1, starting with deformations that are linear in waterfall fields (cases 1a, 1b and 1c) before we turn to an example of a deformation term cubic in waterfall fields (case 1d).

\subsubsection{Case 1a}
As a first example of a linear deformation we consider the following superpotential:
\begin{equation}W=\kappa S \left( H \bar{H}- M^2\right)+\frac{\zeta}{\Lambda} (\phi \bar{\phi})(H \bar{H})+\delta (H \bar{\phi}).
\label{superpdef1}
\end{equation}
We note that, equivalently, we could have added the deformation term $\delta (\bar{H} \phi)$ to the superpotential.
\\

\noindent  {\bf Phase Transition Dynamics:} The scalar potential derived from Eq.~(\ref{superpdef1}) is given by
$$V=\left|\kappa \left( H \bar{H}- M^2\right)\right|^2+\left|\frac{\zeta}{\Lambda} \bar{\phi} (H\bar{H}) \right|^2+\left|\left(\delta+\frac{\zeta}{\Lambda} \phi \bar{H}\right)H\right|^2+\left|\kappa S \bar{H} +\frac{\zeta}{\Lambda} (\phi \bar{\phi}) \bar{H}+\delta \bar{\phi}\right|^2$$\begin{equation}+\left|\left( \kappa S +\frac{\zeta}{\Lambda} (\phi \bar{\phi}) \right) H\right|^2
+\frac{g^2}{2} \left(|\phi|^2-|\bar{\phi}|^2+|H|^2-|\bar{H}|^2\right)^2.\end{equation}
Due to the deformation term $\delta (H \bar{\phi})$ in the superpotential, the vevs of the waterfall fields are not stabilized at zero anymore during inflation. We assume that $\delta$ is a sufficiently small deformation such that the waterfall fields obtain only small vevs during inflation and the approximate flatness of the inflationary field trajectory is preserved.
We furthermore assume in this section that $S$ gets a mass term from the Kähler potential that stabilizes it at zero.

Since the waterfall fields are now nonzero during inflation, the full $D$-flatness condition has to be considered, i.e.\ $\bar{\phi}$ now has to satisfy
\begin{equation}
    |\bar{\phi}|^2=|\phi|^2+|H|^2-|\bar{H}|^2.
    \label{eq3}
\end{equation}
Using $D$-flatness and the smallness of the waterfall fields during inflation, the potential becomes 
\begin{align} V = &\left|\kappa \left( H \bar{H}- M^2\right)\right|^2+\frac{|\zeta|^2}{\Lambda^2}|H|^2|\bar{H}|^2 \left(|\phi|^2+|H|^2-|\bar{H}|^2 \right)+\left|\left(\delta+\frac{\zeta}{\Lambda} \phi \bar{H}\right)H\right|^2 \nonumber\\& +\left|\frac{\zeta}{\Lambda} \phi \bar{H}+\delta \right|^2\left(|\phi|^2+|H|^2-|\bar{H}|^2\right)+\frac{|\zeta|^2}{\Lambda^2} |\phi|^2 |H|^2 \left(|\phi|^2+|H|^2-|\bar{H}|^2\right) \nonumber
\\ \simeq &  |\kappa|^2M^4-|\kappa|^2 M^2 H \bar{H}-|\kappa|^2M^2 H^* \bar{H}^*+\frac{|\zeta|^2}{\Lambda^2}|\phi|^4 |H|^2+
\left|\delta+\frac{\zeta}{\Lambda}\bar{H}\phi\right|^2|\phi|^2 \nonumber
\\ &+ 2|\delta|^2|H|^2-|\delta|^2|\bar{H}|^2,\end{align}
where in the last equality we consider only terms up to second order in the waterfall fields.

The inflationary trajectory during inflation can be obtained by assuming that the waterfall fields are sufficiently heavy to track their potential minima, such that we can effectively integrate them out. Using 
\begin{eqnarray}0&=&\frac{\partial V}{\partial \bar{H}}=-|\kappa|^2 M^2 H+\left(\delta^*+\frac{\zeta^*}{\Lambda} \bar{H}^* \phi^*\right) \frac{\zeta}{\Lambda} \phi |\phi|^2-|\delta|^2 \bar{H}^*,\\
0&=&\frac{\partial V}{\partial H}=-|\kappa|^2 M^2 \bar{H}+ \frac{|\zeta|^2}{\Lambda^2} |\phi|^4 H^*+2|\delta|^2 H^*,\end{eqnarray}
we thus obtain 
\begin{equation}H(\phi)=\frac{M^2 |\kappa|^2 \delta^* \frac{\zeta}{\Lambda} |\phi|^2 \phi}{2 |\delta|^4+M^4 |\kappa|^4-|\delta|^2 \frac{|\zeta|^2}{\Lambda^2} |\phi|^4 -\frac{|\zeta|^4}{\Lambda^4} |\phi|^8},\end{equation}
\begin{equation}\bar{H}(\phi)=\frac{\left(2 |\delta|^2+\frac{|\zeta|^2}{\Lambda^2} |\phi|^4\right) \delta \frac{\zeta^*}{\Lambda} |\phi|^2 \phi^*}{2 |\delta|^4+M^4 |\kappa|^4-|\delta|^2 \frac{|\zeta|^2}{\Lambda^2} |\phi|^4 -\frac{|\zeta|^4}{\Lambda^4} |\phi|^8}.\end{equation}
We see that $\text{arg}(\bar{H})\equiv -\text{arg}(H)\equiv \text{arg}(\delta \zeta^* \phi^*)\ (\text{mod} \ 2 \pi)$ and that for $\delta \neq 0$ the waterfall fields are indeed non-zero already during inflation. 

Plugging this into $V$ we obtain the leading order effective inflaton potential\footnote{We note here again that the leading order inflaton potential gets additional contributions, e.g.\ from the Kähler potential, from SUSY breaking effects as well as by loop corrections. We assume that they allow to realise the measured values of the inflationary CMB observables.}
\begin{equation}
V(\phi)\simeq M^4 |\kappa|^2 + |\delta|^2 |\phi|^2 +\frac{2 |\delta|^4 \frac{|\zeta|^2}{\Lambda^2}}{2|\delta|^4+M^4|\kappa|^4} |\phi|^6.
\end{equation}
It depends only on the modulus of $\phi$ and thus there is a degeneracy in $\text{arg}(\phi)$ and $\text{arg}(\bar{\phi})$. However, after the universe has inflated for many e-folds, there will be one fixed value of the angles of $\phi$ and $\bar{\phi}$ in the whole observable universe. Since the waterfall fields depend on $\phi$, their phases are also fixed by the value of $\theta = \text{arg}(\phi)$ during inflation.

At some point, the slope of the field trajectory will become steep and inflation ends. Nevertheless, the phases remain fixed until the fields finally reach the global, SUSY-conserving minimum where 
\begin{equation}\langle \bar{\phi} \rangle=0, \quad \langle \phi \rangle=-\frac{\delta \sqrt{\frac{|\zeta|}{\Lambda}}}{\frac{\zeta}{\Lambda}}\frac{1}{\left(M^4 \frac{|\zeta|^2}{\Lambda^2}+|\delta|^2\right)^{1/4}} e^{i \theta},\nonumber\end{equation}
\begin{equation}\langle H \rangle = \frac{M^2 \sqrt{\frac{|\zeta|}{\Lambda}}}{\left(M^4 \frac{|\zeta|^2}{\Lambda^2}+|\delta|^2\right)^{1/4}} e^{i \theta},\quad \langle \bar{H} \rangle=\frac{\left(M^4 \frac{|\zeta|^2}{\Lambda^2}+|\delta|^2\right)^{1/4}}{ \sqrt{\frac{|\zeta|}{\Lambda}}} e^{-i\theta}.\end{equation}
\\

\noindent  {\bf Effect on Topological Defect Formation:}
As described above, the phases of the waterfall fields are fixed by the value of $\text{arg}(\phi)$ during inflation. The waterfall fields never encounter a tachyonic instability and thus move to the same unique vacuum everywhere in the whole observable universe. This implies that no topological defects are produced. 
\\

\subsubsection{Case 1b}
Another example for a linear deformation is given by the superpotential
\begin{equation}W=\kappa S (H\bar{H}-M^2)+\frac{\zeta}{\Lambda}(\phi \bar{\phi})(H \bar{H})+\frac{\delta}{\Lambda} (\phi \bar{\phi})(\phi \bar{H}).\end{equation}
We note that alternatively we could have added the term $\frac{\delta}{\Lambda} (\phi \bar{\phi})(\bar{\phi} H)$, having the same effect. 
\\

\noindent  {\bf Phase Transition Dynamics:}  With the additional terms the scalar potential now reads 
$$V=|\kappa (H\bar{H}-M^2)|^2 +\left| \left(\frac{\zeta}{\Lambda} H +\frac{\delta}{\Lambda} \phi\right) \bar{H} \phi\right|^2+\left|\left(\frac{\zeta}{\Lambda} H +2 \frac{\delta}{\Lambda} \phi \right) \bar{H} \bar{\phi}\right|^2$$\begin{equation}+\left| \left(\kappa S + \frac{\zeta}{\Lambda} \phi \bar{\phi}\right) \bar{H}\right|^2 +\left| \kappa S H +\frac{\zeta}{\Lambda} (\bar{\phi} \phi) H + \frac{\delta}{\Lambda} (\bar{\phi} \phi) \phi \right|^2 + \frac{g^2}{2} \left(|\phi|^2-|\bar{\phi}|^2+|H|^2-|\bar{H}|^2\right)^2. \end{equation}
Again, we assume that $\delta$ is sufficiently small such that due to the deformation, the waterfall fields only obtain small vevs during inflation. The singlet $S$ is stabilized at zero and the $D$-flatness condition is given in Eq.~(\ref{eq3}). Using this and neglecting terms higher than second order in the waterfall fields yields 
$$V\simeq |\kappa|^2 M^4-|\kappa|^2 M^2 H^* \bar{H}^*-|\kappa|^2 M^2 H \bar{H}+ 4 \frac{|\delta|^2}{\Lambda^2} |\phi|^4 |\bar{H}|^2+\frac{|\zeta|^2}{\Lambda^2} |\phi|^4 |\bar{H}|^2$$\begin{equation}+\frac{|\delta|^2}{\Lambda^2} |\phi|^4 |H|^2+\frac{|\zeta|^2}{\Lambda^2} |\phi|^4 |H|^2+\frac{|\delta|^2}{\Lambda^2} |\phi|^6+\frac{\zeta \delta^*}{\Lambda^2} \phi^* |\phi|^4 H + \frac{\zeta^* \delta}{\Lambda^2} \phi |\phi|^4 H^*.\end{equation}
Integrating out the waterfall fields during inflation gives
\begin{equation}H(\phi)=\frac{\frac{\delta \zeta^*}{\Lambda^2} \left(4 \frac{|\delta|^2}{\Lambda^2}+\frac{|\zeta|^2}{\Lambda^2} \right) |\phi|^8 \phi}{M^4 |\kappa|^4- \left(4 \frac{|\delta|^4}{\Lambda^4}+5\frac{|\delta|^2|\zeta|^2}{\Lambda^4} +\frac{|\zeta|^4}{\Lambda^4} \right)|\phi|^8},\end{equation}
\begin{equation} \bar{H}(\phi)=\frac{M^2 |\kappa|^2 \frac{\delta^* \zeta}{\Lambda^2} |\phi|^4 \phi^*}{M^4 |\kappa|^4- \left(4 \frac{|\delta|^4}{\Lambda^4}+5\frac{|\delta|^2|\zeta|^2}{\Lambda^4} +\frac{|\zeta|^4}{\Lambda^4} \right)|\phi|^8}.\end{equation}
For $\delta\neq 0$ the waterfall fields are again non-zero during inflation and $\text{arg}(\bar{H})\equiv -\text{arg}(H)\ (\text{mod} \ 2 \pi)$. 
This leads to the leading order effective inflaton potential
\begin{equation}
V\simeq M^4 |\kappa|^2+ \frac{|\delta|^2}{\Lambda^2} |\phi|^6.
\end{equation}
After inflation, the fields reach the global minimum 
\begin{equation}\langle \phi \rangle= \langle \bar{\phi} \rangle=0, \quad \langle H \rangle = M e^{i \theta}, \quad \langle \bar{H} \rangle= M e^{- i \theta},\end{equation}
where the angle $\theta$ has a fixed value everywhere in the whole observable universe.
\\

\noindent  {\bf Effect on Topological Defect Formation:} 
With the fields following the same trajectory, leading to a unique vacuum everywhere in the whole observable universe, no topological defects are produced.

\subsubsection{Case 1c}
Another possibility for a linear deformation is given by
\begin{equation}
W=\kappa S( H \bar{H}-M^2) +\frac{\zeta}{\Lambda} (\phi \bar{\phi})(H \bar{H})+\delta S H \bar{\phi},
\end{equation}
or equivalently by the deformation term $\delta S \bar{H} \phi$.
\\

\noindent  {\bf Phase Transition Dynamics:} The above superpotential yields the scalar potential
$$V=\left|\kappa (H \bar{H}-M^2) +\delta H\bar{\phi}\right|^2 + \left|\frac{\zeta}{\Lambda} \bar{\phi} H \bar{H}\right|^2+\left|\frac{\zeta}{\Lambda} \phi H \bar{H}+ \delta S H\right|^2$$\begin{equation}+\left|\kappa S \bar{H}+\delta S \bar{\phi}+\frac{\zeta}{\Lambda} (\phi \bar{\phi}) \bar{H}\right|^2+\left|\kappa S H + \frac{\zeta}{\Lambda} \phi \bar{\phi} H\right|^2+\frac{g^2}{2}\left(|\phi|^2-|\bar{\phi}|^2+|H|^2-|\bar{H}|^2\right)^2. \end{equation}
We assume that the singlet field is stabilized at zero. The $D$-flatness condition is given by
\begin{equation}|\phi|^2=|\bar{\phi}|^2-|H|^2+|\bar{H}|^2,\end{equation}
yielding the following potential during inflation:
$$V\simeq M^4 |\kappa|^2-|\kappa|^2 M^2 \bar{H}^* H^*-|\kappa|^2 M^2 \bar{H} H- M^2 \kappa \delta^* \bar{\phi}^* H^*-M^2 \kappa^* \delta \bar{\phi} H $$\begin{equation}+|\delta|^2 |\bar{\phi}|^2 |H|^2+\frac{|\zeta|^2}{\Lambda^2} |\bar{\phi}|^2 |\bar{H}|^2+\frac{|\zeta|^2}{\Lambda^2} |\bar{\phi}|^2 |H|^2.\end{equation}
Integrating out the waterfall fields gives
\begin{equation}H(\bar{\phi})=\frac{-M^2 \kappa \delta^* \frac{|\zeta|^2}{\Lambda^2} |\bar{\phi}|^4 \bar{\phi}^*}{M^4 |\kappa|^4-\frac{|\zeta|^2}{\Lambda^2} |\bar{\phi}|^6 \left(|\delta|^2+\frac{|\zeta|^2}{\Lambda^2} |\bar{\phi}|^2\right)},\end{equation}
\begin{equation}\bar{H}(\bar{\phi})=\frac{-M^4|\kappa|^2 \kappa^* \delta \bar{\phi}}{M^4 |\kappa|^4-\frac{|\zeta|^2}{\Lambda^2} |\bar{\phi}|^6 \left(|\delta|^2+\frac{|\zeta|^2}{\Lambda^2} |\bar{\phi}|^2\right)},\end{equation}
which leads to the leading order effective inflaton potential 
\begin{equation}V\simeq M^4 |\kappa|^2+ \frac{|\delta|^2 \frac{|\zeta|^2}{\Lambda^2}}{|\kappa|^4} 
(|\delta|^2 +|\kappa|^2) |\bar{\phi}|^6.\end{equation}
After inflation, the global minimum is given by
\begin{equation}
\langle \phi \rangle =\langle \bar{\phi} \rangle=0, \quad \langle H\rangle =M e^{i\theta}, \quad \langle \bar{H} \rangle= M e^{-i\theta},
\end{equation}
with $\theta$ being the same everywhere in the observable universe.
\\

\noindent  {\bf Effect on Topological Defect Formation:} As in the cases 1a and 1b, since the waterfall fields follow the same trajectory leading to the same vacuum everywhere in the whole observable universe, no topological defects are produced.

We note that beyond the examples shown above (cases 1a, 1b and 1c) one could also consider more suppressed linear terms like $\frac{\delta}{\Lambda^2} S (\phi \bar{\phi}) (\phi \bar{H})$, motivated for instance by the desire to explain the smallness of the deformation from a suppressed effective operator. As long as the deformation results in non-zero $\langle H \rangle,\langle \bar{H} \rangle$ and a unique field trajectory that leads to the fields ending up in the same vacuum everywhere in the observable universe, the conclusion on the absence of topological defect production remains unchanged.

\subsubsection{Case 1d}
As an example for a deformation term cubic in the waterfall fields, we consider the superpotential
\begin{equation}W=\kappa S (H \bar{H}-M^2) +\frac{\zeta}{\Lambda} (\phi \bar{\phi}) (H \bar{H})+ \frac{\delta}{\Lambda} (\phi \bar{H})(H \bar{H}).\end{equation}
We note that, equivalently, we could have added the term $\frac{\delta}{\Lambda} (\bar{\phi} H)(H \bar{H})$.
\\

\noindent  {\bf Phase Transition Dynamics:} 
Including the cubic deformation term, the scalar potential is given by
$$V=|\kappa (H \bar{H}-M^2)|^2+\left|\frac{\zeta}{\Lambda} \bar{\phi}+\frac{\delta}{\Lambda} \bar{H}\right|^2 |H \bar{H}|^2+\left|\frac{\zeta}{\Lambda} \phi H \bar{H}\right|^2$$\begin{equation}+\left|\kappa S  + \frac{\zeta}{\Lambda} \phi \bar{\phi} + \frac{\delta}{\Lambda} \phi \bar{H}\right|^2|\bar{H}|^2+\left|\kappa S  + \frac{\zeta}{\Lambda} \phi \bar{\phi} + 2\frac{\delta}{\Lambda} \phi \bar{H}\right|^2|H|^2+\frac{g^2}{2}\left(|\phi|^2-|\bar{\phi}|^2+|H|^2-|\bar{H}|^2\right)^2 .\end{equation}
The singlet field is stabilized at zero and the $D$-flatness condition is given by Eq.~(\ref{eq3}). 

In contrast to the liner deformation cases, with the cubic deformation the waterfall fields still have zero vevs $\langle H \rangle=\langle \bar{H} \rangle=0$ during inflation. This means that the squared waterfall masses during inflation (when the fields are expanded around zero waterfall field vevs) and the critical value of the waterfall fields are the same as in the undeformed case 1. As discussed there, the two eigenvalues $m^2_{3,4}$ of $\mathbf{m}^2_\mathcal{H}$ get negative when the inflaton field values drop below a critical point. 

However, in contrast to the discussion there, the cubic term deforms the potential such that already before the critical point is reached, another local minimum arises in the potential, which is energetically lower than the one at $\langle H \rangle=\langle \bar{H} \rangle=0$. Above the critical point, however, there is a potential barrier separating the two minima. At the critical point, the potential barrier vanishes and the point $\langle H \rangle=\langle \bar{H} \rangle=0$ becomes a saddle point. From the unstable saddle point, and even a bit earlier by tunneling, (quantum) fluctuations of the waterfall fields can allow them to move to the energetically favoured minimum. On the other hand, below the critical point, where $m^2_{3,4}<0$, $\langle H \rangle=\langle \bar{H} \rangle=0$ becomes a maximum and in regions where the waterfall fields are located at this point they can move in any direction of the two-dimensional field space. 

Compared to case 1, the potential is tilted, such that there is one unique minimum (which corresponds to the one that formed already before the critical point). The global minimum is reached when
\begin{equation} \langle \phi\rangle =0, \quad \langle \bar{\phi}\rangle= -\frac{\delta}{\zeta} \frac{M}{\left(1+\frac{|\delta|^2}{|\zeta|^2}\right)^{1/4}} e^{i \theta},\nonumber\end{equation} 
\begin{equation} \langle H\rangle=M \left(1+\frac{|\delta|^2}{|\zeta|^2}\right)^{1/4} e^{-i \theta}, \quad \langle \bar{H}\rangle=  \frac{M}{\left(1+\frac{|\delta|^2}{|\zeta|^2}\right)^{1/4}} e^{i \theta}.\end{equation}
In this minimum, we have $V = 0$ and SUSY is conserved. 
\\

\noindent  {\bf Effect on Topological Defect Formation:} 
When the critical point is reached and the waterfall field potential features a saddle point, in most regions of the universe the waterfall field will move to the new energetically favoured minimum. However, it is possible that in some smaller part of the universe, the waterfall fields remain at zero field values. From this maximum, they can roll in any direction of the two-dimensional field space, leading to the formation of cosmic strings. Due to the strong bias towards the energetically favoured minimum, we expect that this happens in smaller regions of the universe, such that string loops of reduced size are produced. Furthermore, since the potential is tilted, on one side of the cosmic strings the potential energy is higher than on the other side, which implies that the string loops form boundaries of temporary domain walls. 

It is energetically favoured to reduce the size of these domain walls, which accelerates the shrinking of the cosmic string loops. Depending on various parameters, such as the size of the cubic deformation term and the details of the inflaton potential, the unstable system of domain walls bounded by strings can effectively disappear. A more detailed dedicated study of this case, which would involve a careful simulation of the quantum fluctuations of the waterfall fields, is beyond the scope of the present work but would be highly desirable.    

Finally, we note that one could also consider cubic waterfall terms from more suppressed higher dimensional operators in the superpotential, for example terms like $\frac{\delta}{\Lambda^2} S (H \bar{H}) (H \bar{\phi})$, to explain the smallness of the deformation. As long as the deformation induced a small cubic term in the waterfall potential, the results are qualitatively unchanged.

\subsection{Effect of a Fayet-Iliopoulos Term}
In general, having a supersymmetric Abelian gauge theory, one can include a Fayet-Iliopoulos term in the Lagrangian, i.e.\
\begin{equation}\mathcal{L}_{FI}=-2 \rchi [V]_D=-\rchi D,\end{equation}
where $V$ is the vector superfield corresponding to the $U(1)$ gauge symmetry and $\rchi$ is real. 
Therefore, the scalar potential becomes
\begin{equation}V=\rchi-\frac{1}{2} D^2-g D \sum_i q_i |\phi_i|^2,\end{equation}
where $\phi_i$ is the scalar field contained in the $i$-th chiral superfield and $q_i$ is its $U(1)$ charge. The equation of motion gives
\begin{equation}D=\rchi-g\sum_i q_i |\phi_i|^2.\end{equation}
Thus, the $D$-term potential
is given by
\begin{equation}
V_D=\frac{1}{2}\left(\rchi-g\left(|\phi|^2-|\bar{\phi}|^2+|H|^2-|\bar{H}|^2\right)\right)^2.
\end{equation}
The Fayet-Iliopoulos term modifies the $D$-flatness condition, but leaves the $F$-term part of the scalar potential unchanged. 
More precisely, during inflation we get for the leading order inflaton potential, using $\langle S\rangle =\langle H\rangle=\langle \bar{H}\rangle=0$, 
\begin{equation}
V_{\text{inf}}=|\kappa|^2 M^4 +\frac{1}{2}\left(\rchi-g \left(|\phi|^2-|\bar{\phi}|^2\right)\right)^2.
\end{equation}
We see that the $D$-term pushes the inflaton fields into a $D$-flat direction, where it now holds that 
\begin{equation}
|\phi|^2=|\bar{\phi}|^2+\frac{\rchi}{g}.
\label{eq:D-termcondFI}
\end{equation}

Considering case 1 as an example and setting $\rchi>0$ for definiteness, the waterfall field during inflation is given by
\begin{equation}m^2_{1,2}=\frac{|\zeta|^2}{\Lambda^2} |\bar{\phi}|^2\left( |\bar{\phi}|^2+\frac{\chi}{g}\right)^2+M^2|\kappa|^2, \quad m^2_{3,4}=\frac{|\zeta|^2}{\Lambda^2} |\bar{\phi}|^2\left( |\bar{\phi}|^2+\frac{\chi}{g}\right)^2-M^2|\kappa|^2.\end{equation}
The critical value below which $m_{3,4}^2<0$ is given by
\begin{equation}
|\bar{\phi}_{\text{crit}}|^2=-\frac{\rchi}{2g}+\frac{1}{2} \frac{\Lambda}{|\zeta|}\sqrt{\frac{|\zeta|^2}{\Lambda^2}\frac{\rchi^2}{g^2}+4M^2|\kappa|^2}.
\end{equation}
Since $m_{1,2}^2>0$ for any $|\bar{\phi}|$ it follows that, at the critical point, $\langle \bar{H} \rangle=\langle H^* \rangle$ (i.e.\ $\text{arg}(\langle \bar{H}\rangle )\equiv-\text{arg}(\langle H\rangle ) (\text{mod } 2\pi)$ and $|\langle \bar{H}\rangle|=|\langle H\rangle|$). Below the critical point, there is still a fixed relation between $\langle \bar{H} \rangle$ and $\langle H^* \rangle$ which ensures that $\text{arg}(\langle \bar{H}\rangle )\equiv-\text{arg}(\langle H\rangle ) (\text{mod } 2\pi)$.   

Plugging this into the potential, we find that the potential only depends on $|H|$ and is independent of $\text{arg}(H)$, i.e.\ there is a degeneracy in $\text{arg}(H)$ and thus, as soon as the waterfall fields become tachyonic, cosmic strings form. We thus find that when we include a Fayet-Iliopoulos term in case 1, the conclusions on topological defect production are unchanged.  

Finally, the global SUSY conserving minimum of the potential is reached when $\langle S \rangle =\langle \phi\rangle= \langle \bar{\phi}\rangle=0$,  $|H|^2=|\bar{H}|^2+\frac{\rchi}{g}$ and $H\bar{H}=M^2$, from which it follows that
\begin{equation}
\langle H\rangle=\frac{\sqrt{\frac{\rchi}{g}+\sqrt{\frac{\rchi^2}{g^2}+4M^4}}}{\sqrt{2}}e^{i \theta(\mathbf{x})} ,\quad \langle \bar{H} \rangle=\frac{\sqrt{-\frac{\rchi}{g}+\sqrt{\frac{\rchi^2}{g^2}+4M^4}}}{\sqrt{2}}e^{-i \theta(\mathbf{x})},
\end{equation}
where the value of the angle $\theta(\mathbf{x})$ varies for different points $\mathbf{x}$ in space.

The other cases can be discussed analogously. The main effect of including the Fayet-Iliopoulos term is again that the D-term condition changes to the one in Eq.~(\ref{eq:D-termcondFI}). 
Regarding in particular the cases 2 and 3, by inspecting the mass eigenvalues and eigenvectors of the waterfall fields one can show that the qualitative statements on the critical points and the phase transition dynamics remain unchanged, and thus also the conclusions on topological defect production.

\section{Applicability to SO(10) Embeddings and Generalisation of our Results}\label{sec:generalisation}

Tribrid inflation offers attractive possibilities for realising inflation in close contact to particle physics models. On the one hand, the inflaton can be the scalar component of a matter superfield, or a $D$-flat direction of matter fields. On the other hand, the phase transition ending inflation can be part of the spontaneous breaking of a larger gauge group to $G_{SM} = SU(3)_C\times SU(2)_L\times U(1)_Y$, the one of the Standards Model (SM) of elementary particles. 
The results of this work are directly applicable to models where inflation is associated to the last step of symmetry breaking where $G_{SM} \times U(1)$ spontaneously breaks to $G_{SM}$.

This includes various possibilities, for instance left-right symmetric extensions of the SM, 
Pati-Salam models or SO(10) Grand Unified Theories (GUTs) broken to $G_{SM}$ in multiple steps. We note that our results also apply to cases where one arrives, before the last step of symmetry breaking, at a group with two $U(1)$ factors, which then break to $G_{SM}$ after/during inflation, for example $SO(10)$ broken first to $\mathcal{G}_{3211}=SU(3)_C\times SU(2)_L\times U(1)_R \times U(1)_{B-L}$ and then to the gauge group of the SM. 
The reason here is that with the hypercharge generator satisfying $Y=\tfrac{1}{2}(B-L)+T_R^3$, where $T_{R}^3$ denotes the third generator of $SU(2)_R$, one can rewrite the $U(1)_R \times U(1)_{B-L}$ part as $U(1)_Y \times U(1)_{X}$, with $U(1)_{X}$, generated by the generator $X$ orthogonal $Y$, broken in the last step of SO(10) breaking. 

Let us discuss in a bit more detail how to proceed for the SO(10)-embedding mentioned above. With $F_\alpha\equiv \mathbf{16}_\alpha$ ($\alpha=1,\ldots, 4$) and $\bar{F}\equiv \overline{\mathbf{16}}$ denoting matter representations\footnote{As discussed in \cite{Antusch:2010va}, this leads to the three light generations of charged fermions of the SM, a vector-like heavy generation and five singlet states (right-handed neutrinos).} and the waterfall fields $H\equiv \mathbf{16}$ and $\bar{H}\equiv \overline{\mathbf{16}}$ (which break $\mathcal{G}_{3211} \to G_{SM}$ after inflation) at the SO(10)-level, the superpotential (without deformations) may contain the following terms: 
\begin{eqnarray}
W_{\text{Tribrid}}&=&\kappa S\left(H\bar{H}-M^2\right)+\frac{\zeta_\alpha}{\Lambda} (\bar{F} F_\alpha)(H\bar{H})+\frac{\tilde{\zeta}_\alpha}{\Lambda} (\bar{F} H)(F_\alpha\bar{H}) \nonumber \\
&&+\frac{\lambda_{\alpha \beta}}{\Lambda} (\bar{H} F_\alpha)(\bar{H} F_\beta)+\frac{\gamma}{\Lambda} (\bar{F} H)(\bar{F} H),
\end{eqnarray}
Including only the terms of the first line correspond to case 1, only the terms proportional to $\kappa,\lambda,\gamma$ to case 2, and all terms listed above to case 3. In addition, there may be extra terms that lead to deformations of the potential.
After performing the breaking of $SO(10)\to\mathcal{G}_{3211}$ (and formulating the model in terms of $G_{SM} \times U(1)_\mathrm{B-L}$), one can apply the results of the previous section. 

We note that while our analysis in section 3 may serve as a template, such models with larger gauge groups feature a larger number of matter field as well as waterfall field degrees of freedom. For example, in the above-described SO(10) example on has to choose the field direction for inflation out of the various matter field components, which for instance could be formed by the right-handed sneutrinos, satisfying the $D$-flatness condition $\sum_{\alpha=1}^4 |\nu_{\alpha}^c|^2=|\bar{\nu}^c|^2$. Similarly, one has to analyse which of the waterfall field directions become dynamical and finally break the symmetry, which in the discussed example is again the right-handed sneutrino direction of $H, \bar H$.

Models with multi-step breaking, where in the first steps monopoles are generated (such as the above-mentioned SO(10) scenario), are of particular interest with respect to the recent PTA results, hinting at 
a stochastic gravitational wave (GW) background at nanohertz frequencies \cite{NANOGrav:2023gor,EPTA:2023fyk,Xu:2023wog,Reardon:2023gzh}. 
In such scenarios the cosmic strings can be meta-stable, since they can decay via producing monopole-antimonopole pairs, and meta-stable cosmic strings are among the best fitting explanations for the observed stochastic GW background \cite{NANOGrav:2023hvm,EPTA:2023xxk}.   
Our results show which type of Tribrid inflation model variants produce cosmic strings that, when embedded into scenarios where they are metastable, can lead to a possible explanation of the recent PTA results.

\begin{table}
    \centering
\begin{tabular}{ c |c |c |c |c | c| c| c }
 Case & 1 & 2& 3& 1a& 1b& 1c& 1d\\[0.5ex]
  \hline\rule{0pt}{1.05\normalbaselineskip}
 Defects if $\langle S\rangle =0$ & CS & CS& CS $+$ DW$^*$ &/ & /&/ & /$^{**}$ \\[0.5ex]
  \hline\rule{0pt}{1.05\normalbaselineskip}
 Defects if $\langle S\rangle \neq 0$ & CS & CS $+$ DW$^*$ & CS $+$ DW$^*$ \\[0.5ex]
 \end{tabular}
\caption{Summary of topological defect formation after the classes of Tribrid inflation from table \ref{Tab:1}, distinguishing in addition between the scenarios where $\langle S \rangle =0$ and $\langle S \rangle \neq 0$. CS denotes cosmic strings and DW domain walls. The star indicates that when domain walls appear, they are only temporary and disappear when the inflaton fields settle at zero vevs.  
The doublestar means that this case needs further investigation, as discussed in the main text. }
\label{summary}
\end{table}

\section{Summary and Conclusions}
In this paper, we have investigated the formation of topological defects in classes of Tribrid inflation models associated with the breaking of a gauge symmetry $G = U(1)$. Similar to hybrid inflation, the end of inflation in Tribrid Inflation is reached when a ``waterfall field'', which was stabilized during inflation at a field value where the scalar potential features a large vacuum energy, starts rapidly rolling towards its minimum where the symmetry group $G$ is spontaneously broken. 

However, in contrast to standard supersymmetric Hybrid inflation, where the inflaton is a gauge singlet, the inflaton in Tribrid inflation can be a gauge non-singlet, which, via its vacuum expectation value, already breaks the gauge symmetry during inflation. This raises the question whether topological defects can form after inflation in this type of models, and if so, which types of defects are generated. To systematically address this question, we have classified possible terms in the Tribrid inflation superpotential and then analysed selected model variants. An overview of the discussed cases is given in table \ref{Tab:1}, and a summary of our results can be found in table \ref{summary}.

The first three cases cover the different possibilities for the terms that stabilise the waterfall fields during inflation. We find that in all three cases, in the absence of further deformation terms, cosmic strings are produced, while in case 3 and in case 2 with $\langle S \rangle \neq 0$ also temporary domain walls are generated, which however disappear when the inflaton fields settle at zero vevs.  
It is important to note that in order to arrive at the correct conclusion one has to carefully study the dynamics of the waterfall transition. As we highlighted in section \ref{sec:symmargfailure}, evaluating topological defect formation solely on symmetry arguments can be misleading, and in fact fails for case 3 (and case 2 with $\langle S \rangle \neq 0$). 

We then turned to cases 1a, 1b 1c and 1d, where we have extended case 1 by additional terms that deform the waterfall potential as long as the inflaton vevs are non-vanishing. We showed that such terms can suppress the formation of topological defects. Furthermore, the addition of a Fayet-Iliopoulos term to the cases 1, 2 and 3 was shown not to change the qualitative results on topological defect production. Finally, we have also discussed how our results can be used to analyse tribrid inflation associated with the final step of multi-stage $SO(10)$ breaking, where the cosmic strings can be metastable and provide a promising explanation of the recent PTA results hinting at a stochastic gravitational wave background at nanohertz frequencies. 

\section*{Acknowledgments}
K.T. acknowledges financial support from the Slovenian Research Agency (research core funding No. P1-0035 and PR-12830).

%\addcontentsline{toc}{section}{References}


\begin{thebibliography}{99}

%\cite{Guth:1980zm}
\bibitem{Guth:1980zm}
A.~H.~Guth,
%``The Inflationary Universe: A Possible Solution to the Horizon and Flatness Problems,''
Phys. Rev. D \textbf{23} (1981), 347-356
doi:10.1103/PhysRevD.23.347
%10118 citations counted in INSPIRE as of 17 Jun 2024

%\cite{Albrecht:1982wi}
\bibitem{Albrecht:1982wi}
A.~Albrecht and P.~J.~Steinhardt,
%``Cosmology for Grand Unified Theories with Radiatively Induced Symmetry Breaking,''
Phys. Rev. Lett. \textbf{48} (1982), 1220-1223
doi:10.1103/PhysRevLett.48.1220
%5215 citations counted in INSPIRE as of 17 Jun 2024

%\cite{Linde:1981mu}
\bibitem{Linde:1981mu}
A.~D.~Linde,
%``A New Inflationary Universe Scenario: A Possible Solution of the Horizon, Flatness, Homogeneity, Isotropy and Primordial Monopole Problems,''
Phys. Lett. B \textbf{108} (1982), 389-393
doi:10.1016/0370-2693(82)91219-9
%6423 citations counted in INSPIRE as of 17 Jun 2024

%\cite{Linde:1983gd}
\bibitem{Linde:1983gd}
A.~D.~Linde,
%``Chaotic Inflation,''
Phys. Lett. B \textbf{129} (1983), 177-181
doi:10.1016/0370-2693(83)90837-7
%3507 citations counted in INSPIRE as of 14 Jun 2024

%\cite{Kibble:1976sj}
\bibitem{Kibble:1976sj}
T.~W.~B.~Kibble,
%``Topology of Cosmic Domains and Strings,''
J. Phys. A \textbf{9} (1976), 1387-1398
doi:10.1088/0305-4470/9/8/029
%3257 citations counted in INSPIRE as of 15 Jun 2024

%\cite{Preskill:1979zi}
\bibitem{Preskill:1979zi}
J.~Preskill,
%``Cosmological Production of Superheavy Magnetic Monopoles,''
Phys. Rev. Lett. \textbf{43} (1979), 1365
doi:10.1103/PhysRevLett.43.1365
%831 citations counted in INSPIRE as of 10 Jun 2024

%\cite{Hindmarsh:1994re}
\bibitem{Hindmarsh:1994re}
M.~B.~Hindmarsh and T.~W.~B.~Kibble,
%``Cosmic strings,''
Rept. Prog. Phys. \textbf{58} (1995), 477-562
doi:10.1088/0034-4885/58/5/001
[arXiv:hep-ph/9411342 [hep-ph]].
%1134 citations counted in INSPIRE as of 12 Jun 2024

%\cite{Linde:1993cn}
\bibitem{Linde:1993cn}
A.~D.~Linde,
%``Hybrid inflation,''
Phys. Rev. D \textbf{49} (1994), 748-754
doi:10.1103/PhysRevD.49.748
[arXiv:astro-ph/9307002 [astro-ph]].
%1444 citations counted in INSPIRE as of 13 Jun 2024

%\cite{Copeland:1994vg}
\bibitem{Copeland:1994vg}
E.~J.~Copeland, A.~R.~Liddle, D.~H.~Lyth, E.~D.~Stewart and D.~Wands,
%``False vacuum inflation with Einstein gravity,''
Phys. Rev. D \textbf{49} (1994), 6410-6433
doi:10.1103/PhysRevD.49.6410
[arXiv:astro-ph/9401011 [astro-ph]].
%1171 citations counted in INSPIRE as of 13 Jun 2024

%\cite{Dvali:1994ms}
\bibitem{Dvali:1994ms}
G.~R.~Dvali, Q.~Shafi and R.~K.~Schaefer,
%``Large scale structure and supersymmetric inflation without fine tuning,''
Phys. Rev. Lett. \textbf{73} (1994), 1886-1889
doi:10.1103/PhysRevLett.73.1886
[arXiv:hep-ph/9406319 [hep-ph]].
%698 citations counted in INSPIRE as of 29 May 2024

%\cite{Linde:1997sj}
\bibitem{Linde:1997sj}
A.~D.~Linde and A.~Riotto,
%``Hybrid inflation in supergravity,''
Phys. Rev. D \textbf{56} (1997), R1841-R1844
doi:10.1103/PhysRevD.56.R1841
[arXiv:hep-ph/9703209 [hep-ph]].
%356 citations counted in INSPIRE as of 21 May 2024

%\cite{Antusch:2009vg}
\bibitem{Antusch:2009vg}
S.~Antusch, K.~Dutta and P.~M.~Kostka,
%``Tribrid Inflation in Supergravity,''
AIP Conf. Proc. \textbf{1200} (2010) no.1, 1007-1010
doi:10.1063/1.3327524
[arXiv:0908.1694 [hep-ph]].
%20 citations counted in INSPIRE as of 29 Feb 2024

%\cite{Antusch:2008pn}
\bibitem{Antusch:2008pn}
S.~Antusch, M.~Bastero-Gil, K.~Dutta, S.~F.~King and P.~M.~Kostka,
%``Solving the eta-Problem in Hybrid Inflation with Heisenberg Symmetry and Stabilized Modulus,''
JCAP \textbf{01} (2009), 040
doi:10.1088/1475-7516/2009/01/040
[arXiv:0808.2425 [hep-ph]].
%66 citations counted in INSPIRE as of 24 Apr 2024

%\cite{Antusch:2009ef}
\bibitem{Antusch:2009ef}
S.~Antusch, K.~Dutta and P.~M.~Kostka,
%``SUGRA Hybrid Inflation with Shift Symmetry,''
Phys. Lett. B \textbf{677} (2009), 221-225
doi:10.1016/j.physletb.2009.05.043
[arXiv:0902.2934 [hep-ph]].
%38 citations counted in INSPIRE as of 13 Mar 2024

%\cite{Antusch:2004hd}
\bibitem{Antusch:2004hd}
S.~Antusch, M.~Bastero-Gil, S.~F.~King and Q.~Shafi,
%``Sneutrino hybrid inflation in supergravity,''
Phys. Rev. D \textbf{71} (2005), 083519
doi:10.1103/PhysRevD.71.083519
[arXiv:hep-ph/0411298 [hep-ph]].
%67 citations counted in INSPIRE as of 14 May 2024

%\cite{Antusch:2010va}
\bibitem{Antusch:2010va}
S.~Antusch, M.~Bastero-Gil, J.~P.~Baumann, K.~Dutta, S.~F.~King and P.~M.~Kostka,
%``Gauge Non-Singlet Inflation in SUSY GUTs,''
JHEP \textbf{08} (2010), 100
doi:10.1007/JHEP08(2010)100
[arXiv:1003.3233 [hep-ph]].
%50 citations counted in INSPIRE as of 16 May 2024

%\cite{Masoud:2021prr}
\bibitem{Masoud:2021prr}
M.~A.~Masoud, M.~U.~Rehman and Q.~Shafi,
%``Sneutrino tribrid inflation, metastable cosmic strings and gravitational waves,''
JCAP \textbf{11} (2021), 022
doi:10.1088/1475-7516/2021/11/022
[arXiv:2107.09689 [hep-ph]].
%29 citations counted in INSPIRE as of 12 Jun 2024

%\cite{Antusch:2023zjk}
\bibitem{Antusch:2023zjk}
S.~Antusch, K.~Hinze, S.~Saad and J.~Steiner,
%``Singling out SO(10) GUT models using recent PTA results,''
Phys. Rev. D \textbf{108} (2023) no.9, 095053
doi:10.1103/PhysRevD.108.095053
[arXiv:2307.04595 [hep-ph]].
%40 citations counted in INSPIRE as of 15 Jun 2024

%\cite{NANOGrav:2023gor}
\bibitem{NANOGrav:2023gor}
G.~Agazie \textit{et al.} [NANOGrav],
%``The NANOGrav 15 yr Data Set: Evidence for a Gravitational-wave Background,''
Astrophys. J. Lett. \textbf{951} (2023) no.1, L8
doi:10.3847/2041-8213/acdac6
[arXiv:2306.16213 [astro-ph.HE]].
%673 citations counted in INSPIRE as of 17 Jun 2024

%\cite{EPTA:2023fyk}
\bibitem{EPTA:2023fyk}
J.~Antoniadis \textit{et al.} [EPTA and InPTA:],
%``The second data release from the European Pulsar Timing Array - III. Search for gravitational wave signals,''
Astron. Astrophys. \textbf{678} (2023), A50
doi:10.1051/0004-6361/202346844
[arXiv:2306.16214 [astro-ph.HE]].
%526 citations counted in INSPIRE as of 17 Jun 2024

%\cite{Xu:2023wog}
\bibitem{Xu:2023wog}
H.~Xu, S.~Chen, Y.~Guo, J.~Jiang, B.~Wang, J.~Xu, Z.~Xue, R.~N.~Caballero, J.~Yuan and Y.~Xu, \textit{et al.}
%``Searching for the Nano-Hertz Stochastic Gravitational Wave Background with the Chinese Pulsar Timing Array Data Release I,''
Res. Astron. Astrophys. \textbf{23} (2023) no.7, 075024
doi:10.1088/1674-4527/acdfa5
[arXiv:2306.16216 [astro-ph.HE]].
%502 citations counted in INSPIRE as of 17 Jun 2024

%\cite{Reardon:2023gzh}
\bibitem{Reardon:2023gzh}
D.~J.~Reardon, A.~Zic, R.~M.~Shannon, G.~B.~Hobbs, M.~Bailes, V.~Di Marco, A.~Kapur, A.~F.~Rogers, E.~Thrane and J.~Askew, \textit{et al.}
%``Search for an Isotropic Gravitational-wave Background with the Parkes Pulsar Timing Array,''
Astrophys. J. Lett. \textbf{951} (2023) no.1, L6
doi:10.3847/2041-8213/acdd02
[arXiv:2306.16215 [astro-ph.HE]].
%530 citations counted in INSPIRE as of 17 Jun 2024

%\cite{Bastero-Gil:2006zpr}
\bibitem{Bastero-Gil:2006zpr}
M.~Bastero-Gil, S.~F.~King and Q.~Shafi,
%``Supersymmetric Hybrid Inflation with Non-Minimal Kahler potential,''
Phys. Lett. B \textbf{651} (2007), 345-351
doi:10.1016/j.physletb.2006.06.085
[arXiv:hep-ph/0604198 [hep-ph]].
%135 citations counted in INSPIRE as of 16 Apr 2024

%\cite{Rehman:2009nq}
\bibitem{Rehman:2009nq}
M.~U.~Rehman, Q.~Shafi and J.~R.~Wickman,
%``Supersymmetric Hybrid Inflation Redux,''
Phys. Lett. B \textbf{683} (2010), 191-195
doi:10.1016/j.physletb.2009.12.010
[arXiv:0908.3896 [hep-ph]].
%89 citations counted in INSPIRE as of 10 Apr 2024

%\cite{Nakayama:2010xf}
\bibitem{Nakayama:2010xf}
K.~Nakayama, F.~Takahashi and T.~T.~Yanagida,
%``Constraint on the gravitino mass in hybrid inflation,''
JCAP \textbf{12} (2010), 010
doi:10.1088/1475-7516/2010/12/010
[arXiv:1007.5152 [hep-ph]].
%59 citations counted in INSPIRE as of 14 May 2024

%\cite{Antusch:2012jc}
\bibitem{Antusch:2012jc}
S.~Antusch and D.~Nolde,
%``K\"ahler-driven Tribrid Inflation,''
JCAP \textbf{11} (2012), 005
doi:10.1088/1475-7516/2012/11/005
[arXiv:1207.6111 [hep-ph]].
%14 citations counted in INSPIRE as of 04 Apr 2024

%\cite{Buchmuller:2014epa}
\bibitem{Buchmuller:2014epa}
W.~Buchm\"uller, V.~Domcke, K.~Kamada and K.~Schmitz,
%``Hybrid Inflation in the Complex Plane,''
JCAP \textbf{07} (2014), 054
doi:10.1088/1475-7516/2014/07/054
[arXiv:1404.1832 [hep-ph]].
%57 citations counted in INSPIRE as of 06 Jun 2024

%\cite{Schmitz:2018nhb}
\bibitem{Schmitz:2018nhb}
K.~Schmitz and T.~T.~Yanagida,
%``Axion Isocurvature Perturbations in Low-Scale Models of Hybrid Inflation,''
Phys. Rev. D \textbf{98} (2018) no.7, 075003
doi:10.1103/PhysRevD.98.075003
[arXiv:1806.06056 [hep-ph]].
%19 citations counted in INSPIRE as of 13 Jun 2024

%\cite{NANOGrav:2023hvm}
\bibitem{NANOGrav:2023hvm}
A.~Afzal \textit{et al.} [NANOGrav],
%``The NANOGrav 15 yr Data Set: Search for Signals from New Physics,''
Astrophys. J. Lett. \textbf{951} (2023) no.1, L11
doi:10.3847/2041-8213/acdc91
[arXiv:2306.16219 [astro-ph.HE]].
%418 citations counted in INSPIRE as of 17 Jun 2024

%\cite{EPTA:2023xxk}
\bibitem{EPTA:2023xxk}
J.~Antoniadis \textit{et al.} [EPTA and InPTA],
%``The second data release from the European Pulsar Timing Array - IV. Implications for massive black holes, dark matter, and the early Universe,''
Astron. Astrophys. \textbf{685} (2024), A94
doi:10.1051/0004-6361/202347433
[arXiv:2306.16227 [astro-ph.CO]].
%253 citations counted in INSPIRE as of 17 Jun 2024

\end{thebibliography}
\end{document}